%Paper, with date.

 %\font\steptwo=cmbx scaled\magstep2
% \font\stepthree=cmbx scaled\magstep3
 \magnification=\magstep1
\settabs 18 \columns
%paper, date, \b, \q, \r, \ce, ,ve,
%\YB, \UT,
%\hoffset=1.00truein
%\voffset=1.00truein
\hsize=16truecm

\def\btd{\bigtriangledown}

\def\b{\bigskip}
\def\bb{\bigskip\bigskip}

\def\no{\noindent}
\def\r{\rightline}
\def\ce{\centerline}
\def\ve{\vfill\eject}

\def\r{\rightline}

\def\L{{\cal L}}

\def\harr#1#2{\smash{\mathop{\hbox to .25 in{\rightarrowfill}}
 \limits^{\scriptstyle#1}_{\scriptstyle#2}}}

\def\R{{\cal R}} 
\def\V{{\cal V}}

\def\today{\ifcase\month\or January\or February\or March\or April\or
May\or June\or July\or
August\or September\or October\or November\or  December\fi
\space\number\day, \number\year }

\r \today
\bb\bb\bb
%\parindent=0pt

%Ecriture des corps de nombres

\def\Rrm{\hbox{\rm I\hskip -2pt R}}

\def\e{\rm e}
\def\d{\delta}
\def\p{\partial}

\def\sqr#1#2{{\vcenter{\vbox{\hrule height.#2pt
\hbox{\vrule width.#2pt height#2pt \kern#2pt
\vrule width.#2pt}
\hrule height.#2pt}}}}

 \def\1/2{{\scriptstyle{1\over 2}}}
 \def\a/2{{\scriptstyle{3\over 2}}}
 \def\5/2{{\scriptstyle{5\over 2}}}
 \def\7/2{{\scriptstyle{7\over 2}}}
 \def\3/4{{\scriptstyle{3\over 4}}}

\font\steptwo=cmb10 scaled\magstep2
\font\stepthree=cmb10 scaled\magstep4
\magnification=\magstep1

\def\sqr#1#2{{\vcenter{\vbox{\hrule height.#2pt
\hbox{\vrule width.#2pt height#2pt \kern#2pt
\vrule width.#2pt}
\hrule height.#2pt}}}}

\def \r{\rightarrow}
\def\M{{\cal M}}

  {\ce {\steptwo   Thermodynamics with an Action Principle (2nd edition)   }}.

\ce{\bf Heat and gravitation}
\b
  \ce{Christian Fr\o nsdal}
\b
  \ce{\it Physics Department, University of California, Los Angeles CA
  90095-1547 USA}
 \b

\def\sqr#1#2{{\vcenter{\vbox{\hrule height.#2pt
\hbox{\vrule width.#2pt height#2pt \kern#2pt
\vrule width.#2pt}
\hrule height.#2pt}}}}

\def \r{\rightarrow}
\def\M{{\cal M}}

  \no ABSTRACT ~  Some features of hydro- and thermodynamics, as applied to
atmospheres and to stellar structures, are puzzling:  1.  The suggestion,
first made by Laplace, that our atmosphere has an adiabatic temperature
distribution, is confirmed for the lower layers, but the reason why it should be
so is understood only qualitatively.   2.  
Arguments in which a concept of energy plays a role, in the context of
hydro-thermodynamical systems and gravitation, are often 
flawed, and some familiar results concerning the
stability of model stellar structures, first advanced at the end of the
19th century  and repeated in the most modern textbooks, are less than
completely convincing.   3. The
standard treatment of relativistic thermodynamics does not allow for a
systematic treatment of mixtures, such as the mixture of a perfect gas with
radiation. 4. The concept of mass in applications of general relativity to
stellar structure is unsatisfactory. It is proposed that a formulation of
thermodynamics as an action principle may be a suitable approach to adopt for a
new investigation of these matters. 

We formulate thermodynamics
of ideal gases in terms of an action principle
and study the interaction between an ideal gas and the photon gas, or
heat. The action principle provides a hamiltonian functional, not available 
in traditional approaches where familiar
expressions for the energy have no operative meaning. The usual
polytropic atmosphere in an external gravitational field is examined, in order to
determine to what extent it is shaped by radiation.  It is easy to
understand that radiation sustains the atmosphere  (prevents cooling), but the 
temperature profile is largely determined by intrinsic properties of the gas and
it is difficult to interpret it as an effect of radiation. This has led some
people to question whether the rule of uniform temperature as an absolute
condition for equilibrium is valid in the presence of a gravitational
field. An experiment that involves a centrifuge and that has wider
implications in view of the equivalence principle, is proposed, to ascertain the
influence of gravitation on the equilibrium distribution with a very high degree
of precision.   A new formulation of the concept of radiative
equilibrium is proposed.
 
The choice of boundary conditions for radial, stellar stability calculations is
clarified with the help of a properly defined, conserved mass distribution.  

\b
PACS Keywords: Atmosphere, photon gas, action principle.
  \ve
\no{\steptwo I. Introduction}

 The statement that any two thermodynamic systems, each in a state of
equilibrium with a well defined temperature, in thermal equilibrium with each
other, must have the same
temperature, is a central tenet of thermodynamics. A natural extension is
that the temperature in an extended but closed system in a state of
equilibrium  must be uniform, but
there does not seem to be universal agreement on whether this is true in the
presence of gravitational fields. The question comes up in the  investigation of
terrestrial or stellar atmospheres, where the gravitational forces create a
non-uniform density distribution. 

This paper is a study of the polytropic atmosphere. We want to know if (or to
what extent) the polytropic relations are to be attributed to intrinsic
properties of the gas, or to radiation. In this we are guided by   a strong
preference for action principles.

The textbooks present hydrodynamics as the theory of a continuous
distribution of matter,  described in the simplest case by two fields or
distributions: a density field and a velocity field, both defined over
$\Rrm^3$ or a portion thereof.  The role of temperature is often disguised,
assumed to be determined by the density and the pressure.
Classical thermodynamics, on the other hand, is primarily the study of
states of equilibrium, with uniform density and temperature,  and relations
between such states. In this context extremum principles play a pivotal role;
see for example Callen (1960). Texts that deal with  flow of matter or with
temperatures that vary in time or from one point to another are found under the
heading of heat transfer, fluctuations or thermodynamics of irreversible
processes. See for example Stanyukovich (1960),   M\"uller (2007). These studies
rely heavily on conservation laws, but variational principles are rarely
mentioned. Texts that most closely parallell the present work are found under 
radiation hydrodynamics (Castor 2004).

In this introduction we study a simple system from the point of view
of hydrodynamics, on the basis of a well known action principle. The concept
of temperature appears, but not as a dynamical variable. We offer a brief review
of the history of the polytropic atmosphere (Section I.6) and stress the role
of mass.  In
Section II, we extend the action principle to include the temperature as  
an independent field variable.  A lagrangian  
describes the configurations on a single polytrope.  The Euler-Lagrange
equations include the gas law as well as the polytropic relation between
pressure and density. The familiar expression for the internal energy of an
ideal, polytropic gas coincides with the hamiltonian; this appears to be a
significant result. The energy and the pressure of radiation are incorporated in
a natural way  (Section II.6). 

 The theory is mathematically complete in the sense that
no additional input from underlying microscopic physics is needed;
as an example we derive a virial theorem that is proper to the action
principle (Section II.7). But it is 
physically incomplete since the role of convection and
heat flow is not explicitly accounted for. 

To contribute to the debate on the
question of whether an isolated atmosphere in a gravitational field tends to
isothermal equilibrium we study an ideal gas in a centrifugue and invoke the
equivalence principle (Section II.8).

 In Section III we study the effect of
sources of heat that induce transitions between polytropes.   The radiation field
contributes as a source of entropy.   The principal idea behind this work was the hope of understanding, in a
quantitative way, why real atmospheres tend to be polytropic; in this we have
had only limited success.

 In  Section IV we take up the
problem of the stability of polytropic, atmospheric models. A principal
advantage of the method is that it provides us with a hamiltonian, expressed in
terms of the dynamical variables. Some classical stability studies are found
wanting, because of {\it ad hoc} definitions of various energies, and
inapropriate  boundary conditions.
 
 Section V
makes the passage to General Relativity. Section VI has a summary
of conclusions and several proposals for additional work, theoretical as
well as experimental.

\b\b
\ce{\bf I.1. Hydrodynamics}

The textbook introduction to hydrodynamics deals with a density field
$\rho$ and a velocity field $\vec v$ over $\Rrm^3$, subject to two
fundamental equations, the equation of continuity,
$$
\dot \rho + {\rm div}(\rho \vec v) = 0,~~ \dot \rho := {\p\rho\over \p
t},\eqno(1.1)
$$
and the hydrodynamical equation (Bernoulli 1738)
$$
-{\rm grad}~ p = \rho{D\over Dt}\vec v := \rho(\dot{\vec v} + \vec v\cdot
{\rm grad} ~\vec v).\eqno(1.2)
$$
This involves another  field, the scalar field $p$, interpreted as the
local pressure. The theory is incomplete and requires an
additional equation relating $p$ to $\rho$. It is always
assumed that this relation is local, giving $p(x)$ in terms of the
density  at the same point $x$, and instantaneous.

\b\b

\ce{\bf I.2. Laminar flow}

Since we are reluctant to take on  difficult problems of turbulence,
we shall  assume, here and throughout, that the velocity field can be
represented as the gradient of a scalar field,
$$
\vec v = -{\rm grad}~ \Phi.\eqno(1.3)
$$
In this case the hydrodynamical condition is reduced to
$$
   {\rm grad} ~p = \rho~{\rm grad}~ (\dot\Phi - \vec
v^2/2).\eqno(1.4)
$$
To complete this system one needs a relation between the fields $p$ and
$\rho$.

Assume that there is a local
functional
$V[\rho]$ such that
$$
p = \rho V' - V,~~ V' := dV/d\rho.\eqno(1.5)
$$
In this case $dp = \rho~ dV'$ and the equation becomes, if $\rho \neq
0$,
$$
{\rm grad}~ V' = {\rm grad}~(\dot \Phi -   \vec v^2/2)\eqno(1.6)
$$
or
$$V' =  \dot \Phi -  \vec v^2/2 + \lambda,~~ \lambda ~
{\rm constant}.\eqno(1.7)
$$
  The potential $V[\rho]$ is defined by $p$
modulo a linear term, so that the appearance of an arbitrary constant
is natural. It will serve as a
Lagrange multiplier.

{\it The introduction of a velocity potential guarantees the existence of
a first integral of the motion,
a conserved energy functional that will play an important role in the
theory.}
\ve

\ce{\bf I.3. Variational formulation}

Having restricted our scope, to account for laminar flows only,  we have
   reduced the fundamental equations of simple hydrodynamics to the
following two equations,
$$\eqalign{ 
\p_\mu J^\mu& =  0,~~ J^t := \rho,~ \vec J := \rho \vec v, \cr &
V' =  \dot \Phi -  \vec v^2/2 + \lambda,\cr}\eqno(1.8)
$$
together with the defining equations
$$
\vec v = -{\rm grad}~ \Phi,~~ p := \rho V' - V.\eqno(1.9)
$$
It is well known that these equations are the Euler-Lagrange
equations associated with the action (Fetter and Walecka  1980)
$$
A = \int dtd^3x ~{\cal L},~~ {\cal L} =  \rho(\dot\Phi - \vec v^2/2 +
\lambda) - V[\rho].\eqno(1.10)
$$
{\it The value of this last circumstance lies in the fact that the
variational principle is a much better starting point for generalizations,
including the incorporation of symmetries, of special relativity, and the
inclusion of electromagnetic and gravitational interactions. It also gives us a
valid concept of a total energy functional.}

\b 

\ce{\bf I.4. On shell relations }

   The
action (1.10) contains only the fields $\Phi$ and $\rho$, and the
Euler-Lagrange equations define a complete dynamical framework, but only
after specification of the functional $V[\rho]$. The pressure was
defined by Eq.(1.9), $p := \rho V' - V$, and one easily verifies that,
on the trajectory, by virtue of the equations of motion,
$$
p = {\cal L} ~~({\rm on~ shell}).\eqno(1.11)
$$
This fact has been noted, and has led to the
suggestion that the action principle reduce to the minimization of $\int
p$ with respect to variations of $p$ defined by thermodynamics
(Taub 1954), (Bardeen  1970), (Schutz 1970).
   But more is needed, including an off shell action.  After adopting
the action (1.10) it remains  to  relate the choice of the potential $V$
to the thermodynamical properties of the fluid.
\b\b

\ce{\bf I.5. Equation of state and equation  of change}

An ideal gas at equilibrium, with constant temperature, obeys the gas law
$$
p/\rho = \R T.\eqno(1.12)
$$
Pressure and density are in cgs units and 
$$
\R = (1/\mu) \times .8314\times 10^8~erg/K ,
$$
 where $\mu$ is the 
atomic weight. In this paper the validity of the gas law
is assumed to hold, locally at each point of the gas,  under all circumstances, 
including the case that gravitational and electromagnetic fields are present.
Effective values of $\mu$ are
$$
{\rm Atomic ~hydrogen:}~\mu = 1,~~ {\rm Air:}~\mu = 29,~ ~{\rm Sun:}~ \mu = 2.
$$

Equation (1.12) is the only equation that is referred to as an `equation of
state'.
 Other relations, to be introduced next, are `equations of
change', this term taken from Emden's ``Zustands\"anderung", for their
meaning is of an entirely different sort. The most important is the polytropic
relation
$$
p = A\rho^{\gamma'}, ~~ A, \gamma' ~{\rm constant}.\eqno(1.13)
$$
This relation defines a polytropic \underbar {path} or \underbar{polytrope} in 
the $p,v$ diagram ($ v = 1/ \rho $). A polytropic atmosphere is one in which, 
as one moves through the gas, the
variables $\rho$ and $p$ change so as to remain always on the same polytrope,
the temperature being determined by Eq.(1.12) always. Eq.(1.13) is a statement
about the system, not about the gas {\it per se}.

The index of the polytrope is the positive number $n$ defined by
$$
\gamma' =: 1 + {1\over n'}.
$$
Important special cases are
$$
n' = 0,~~ \gamma = \infty,~~\rho = {\rm constant},
$$
$$
\gamma' = C_P/C_V, {\rm ~specific ~entropy~ = constant},
$$
$$
n' = \infty, ~~\gamma' = 1,~~T = {\rm constant}.
$$
The numbers $\gamma,~n$ are defined by
$$
\gamma := C_P/C_V =: 1 + {1\over n}.
$$
The number $n$ is the
adiabatic index of the gas. According to statistal mechanics $2n$ is the number
of degrees of freedom of each molecule in the gas.   That atmospheres tend to be
polytropic is an empirical fact.   

The case that
$\gamma' =
\gamma$ is of a special significance. 
A polytrope  with $\gamma' = \gamma$ is a path of  constant  specific
entropy; changes along such polytropes are reversible and adiabatic; these
polytropes and no others are adiabats.

Fix the constants $A, \gamma' $ and consider an associated  stationary,
polytropic  atmosphere.   Since both (1.12) and (1.13) hold we have
$$
p = {\rm const.} \,\rho^{\gamma'},~~p = 
{\rm const.} \,T^{ {\gamma'\over \gamma'-1} },~~ T = {\rm const.}\,
p^{1-1/\gamma'}.\eqno(1.14)
$$
In any displacement along a polytrope from a point with pressure $p$ and 
temperature $T$, we shall have 
$d\rho/\rho = (1/\gamma')dp/p$, so that an increase in pressure leads to 
an increase in density that is greater for a smaller value of $\gamma'$.  If
a parcel of gas in this atmosphere is pushed down to a region of higher
pressure, by a reversible process, then it will adjust to the ambient pressure.
If $\gamma  > \gamma'$, then it will acquire a density that is
lower than the environment; it will then rise back up; this atmosphere is
stable. But if $\gamma' > \gamma$ then the parcel will be denser than the
environment and it will sink further; this atmosphere
is unstable. Thus we have:
\b
{\it A stable,  polytropic atmosphere must have $\gamma' < \gamma,~~ n' > n$.}
\b
\no Most stable  is the isothermal atmosphere, $\gamma' = 1$.
\b
In hydrodynamics,    the isothermal atmosphere can be given a lagrangian 
treatment by taking
$$
V = \R T \rho \log\rho.\eqno(1.15) 
$$
 We
suppose that the gas is confined to  the section  $z_0<z<z_0 + h$ of a
vertical cylinder with base area ${\cal A}$ and expect the density to fall off
at higher altitudes.  A plausible action density, for a
perfect gas at constant temperature $T$ in a constant gravitational field $\phi
= gz$, $g$ constant, is
$$
{\cal L}[\Phi,\rho] = \rho(\dot\Phi - \vec v^2/2  - gz + \lambda) - {\cal
R}T\rho\log\rho.\eqno(1.16)
$$
We may consider this an isolated system with fixed mass and fixed
extension.

At equilibrium $\dot\Phi = 0, \vec v = 0, \dot \rho = 0$ and the
equation of motion is
$ V' = {\cal R}T(1 + \log\rho) = \lambda - gz,$
hence
$$
\rho(x,y,z) = \e^{ -1+\lambda/\R T}\e^{-gz/\R T},~~ M = {\cal A}{ {\R}
T\over g}\e^{-1 +\lambda/\R T}(1- \e^{-gh/RT})~\e^{-gz_0/\R T}
$$
and after elimination of $\lambda$
$$
\rho = {gM\over {\cal A}{\R} T}{\e^{-g(z-z_0)/\R T}\over 1-
\e^{-gh/RT} },~~ p={gM\over {\cal A}}~{\e^{-g(z-z_0)/\R T}\over 1-
\e^{-gh/RT}}.\eqno(1.17)
$$
There is no difficulty in taking the limit
$h\rightarrow \infty$. The volume becomes infinite but it can be
replaced as
a variable by the parameter $z_0$. This atmosphere is stable; a proof is
presented in Section IV.1. 

The isothermal atmosphere
is usually abandoned in favor of the polytropic atmosphere.

A polytropic gas can be described by the lagrangian (1.10), with
$$
V = \hat  a\rho^{\gamma'}, ~~ \hat a, \gamma' ~{\rm constant}.
$$
Variation with respect to $\rho$  gives
$$
p ={\hat a\over n'}\rho^{\gamma'},~~{1\over n'} = \gamma'-1, 
$$
to be supplemented by the gas law, Eq.(I.12). Among the many
applications  the following are perhaps the most important.
In the case of sound propagation
the gas is initially awakened from equilibrial turpor and then left in
an isolated, frenzied state of oscillating density and pressure,
with the temperature keeping pace in obedience to the gas law.
All three of the relations (1.14) are believed to hold, with $\gamma' =
\gamma$. The oscillations are usually  too rapid for
the heat to disseminate and equalize the temperature,
so that the neglect of heat transfer
may be justified. In applications to the atmospheres one uses the
polytropic equation of change (1.13) and obtains the temperature from 
the gas law. Understanding the resultant temperature gradient in terms of
convection, or as the effect of the heating of the air by solar radiation,
or both, is one of the main issues on which we had hoped to gain some
understanding.

At mechanical equilibrium $\vec v = 0, \dot \rho =
0$ and $\lambda - gz= \hat a\gamma\rho^{1/n}$, hence
$$
  \rho =
({\lambda - gz\over \hat a\gamma})^n.  
$$
Since the density must be
positive one does not fix the volume but assumes that the atmosphere
ends at the point $z_1 = \lambda/g$. Then
$$
 M= {{\cal A}}({g\over
\hat a\gamma })^n\int_{z_0}^{z_1}(z_1-z)^ndz = {{\cal A}h\over
n+1}({gh\over \hat a\gamma})^n. \eqno 
$$
This fixes  $h$ and thus $z_1$
and $ \lambda$. If the atmosphere is an ideal gas then 
 the temperature varies with altitude according
to
$$
{\cal R}T = p/\rho = {\hat a\over n}\rho^{1/n} = g{ z_1-z\over n+1}
\eqno(1.18) 
$$ 
Because the lagrangian does not contain $T$ as a dynamical variable it is
possible to impose this condition by hand.  

 One would not apply this theory
down to the absolute zero of temperature, but even without going to such
extremes it seems risky to be predicting the temperature of
the atmosphere without having made any explicit assumptions about the
absorption or generation of heat that is said to be required to sustain it.
Yet this has been the basis for the phenomenology of stellar structure, as well
as the earth's atmosphere, from the beginning (e.g. Lane 1870, Ritter 1878). 

{\it The success of the isotropic model is notorious, and this success can be
explained in physical terms, but the theory is incomplete since it does not 
account for heat flow, nor convection, both of which are needed to compllete the
picture.}

For air, with
atomic weight 29, $\R = 2.87\times 10^6 ergs/gK$ and $n = 2.5.$  At
sea level,\break$g = 980 cm/sec^2$,  the density  is   $\rho  =1.2
\times 10^{-3}  g/cm^3$, the pressure $ p =1.013 \times 10^6
dyn/cm^2$.  Thus
$$
p/\rho = .844 \times 10^9 cm^2/sec^2, ~~ T = T_0
= 294K,~~ z_1 = 3.014\times 10^6 cm \approx 30km.
$$
and the dry lapse
rate at low altitudes is  $ -T'=294/ z_1= 9.75  K/km.$ The opacity that is
implied by this is mainly due to the presence of $CO_2$ in the atmosphere.
Humidity increases the opacity and decreases the lapse rate by as much as a
factor of 2.

\b\b
\ce{\bf I.6. Historical notes on polytropic atmospheres}

 Observations of reversible transformations of
near-ideal gases,  carried out during the 19th century, can be summarized in
what is sometimes called the laws of Poisson,
$$
\rho\propto T^{n'},~~ p \propto T^{n'+1},~~p \propto \rho^{\gamma'},~~ 
\gamma' = 1 + {1\over n'} ~{\rm constant}.
$$
In the original context all the variables are constant and uniform. 
The exponents as well as the coefficients of proportionality are the same for
all states that are related by reversible transformations. Statistical mechanics
explained this result and confirmed the experimental value $\gamma' = \gamma  =
C_P/C_\V$. As far as can be ascertained, the presence of terrestrial gravitation
and ambient radiation had no effect on these experiments. In a first
extrapolation the same relations were taken to hold locally in
dynamic situations, as in the case of sound propagation.  
The gas is not in equilibrium and the variation of the temperature from point to
point, and with time, is obtained from the gas law.

For the atmosphere of the earth it was at first proposed that the temperature 
would be uniform. However, the existence
of a temperature gradient was soon accepted as an incontrovertible experimental
fact. The first recorded recognition of this, together with an attempt at
explaining the same, may be that of  Carnot, in the paper in which
he created the science of thermodynamics (Carnot 1824). Carnot quotes Laplace: 
``N'est-ce pas au refroidissement  de l'air par la dilatation qu'il faut
attribuer le froid des r\'egions superieures de l'atmosphere? Les raisons
donn\'ees jusqu'ici pour expliquer ce froid  sont tout a fait insuffisantes; on
dit  que l'air des r\'egions elev\'ees, recevant peu de chaleur reflechie par la
terre, et rayonnant lui meme vers les espaces celestes, devait perdre de
calorique, et que c'etait l\'a la cause de son refroidissement; ... " This may be
the first time that  the influence of radiation is
invoked. The temperature gradient is attributed to the greenhouse effect, 
and Laplace was an early skeptic, for he continues   ``...mais cette explication
ce trouve detruite si l'on remarque qu'a
\'egale hauteur le froid regne aussi bien et meme avec plus d'intensit\'e sur
les plaines elev\'ees que sur les sommets des montagnes ou que dans les parties
d'atmosphere \'eloignees du sol." It is not clear that the two explanations are
at odds with each other; Laplace apparently postulates that the atmospheres over
lands at different elevations are related by adiabatic transformations, without
explaining why.

By rejecting the role of radiation as the cause of the temperature gradient, 
Laplace seems to suggest that the same would be observed in an atmosphere
subject to gravitation but totally isolated from radiation, neither exposed to
the radiation coming from the sun nor radiating outwards. As was strongly
emphasized in the later phases of this debate, this would contradict the belief
that the thermal equilibrium of any isolated system, gravitation and other
external forces notwithstanding, is characterized by a uniform temperature. 

In 1862 W. Thomson, in the paper ``On the convective equilibrium of the 
temperature in the atmosphere", defines convective equilibrium with these words
``When all parts of a fluid are freely interchanged and not sensibly influenced
by radiation and conduction, the temperature is said to be in a state of
convective equilibrium." He then goes on to say that an atmosphere that is in
convective equilibrium is a polytrope, and we think that he means an adiabat,
since this is probably implied by the words ``freely interchanged", although 
the value of $\gamma$ is taken from experiment and not from statistical
mechanics. At first sight the clause ``and not sensibly influenced by radiation"
would seem to imply that an isolated atmosphere has a temperature gradient,  but
this conclusion would be premature, as we shall see.

In 1870 H.J. Lane  made the bold assertion that the laws of Poisson may be 
satisfied in the Sun. The terrestrial atmosphere (or part of it) had already been
found  to be well represented by the same relations. Referring to Lane's
paper  Thomson, now lord Kelvin, explains how convective equilibrium
comes about (Thomson 1907). He argues that the atmosphere is not, cannot be, at
rest, and this time radiation plays an essential role. The upper layers loose
heat by radiation and the lower temperature leads to  an increase in density.
This produces a downward current that mixes with a compensating upward drift of
warmer air. This continuing mixing takes place on a time scale that is too short
for adjacent currents to exchange a significant amount of heat by conduction or
radiation, especially since the variations of temperature are very small.   It
is evident that Thomson offers his explanation of the temperature gradient to
account for its absence in an isolated atmosphere,
for he says that, ``an ideal atmosphere, perfectly isolated from absorption as
well as emission of radiation, will, after enough time has passed, reach a state
of uniform temperature, irrespective of the presence of the gravitational
field".   Thomson accepts the mechanism of  Laplace and Carnot, as it is at work
in the real atmosphere, but he goes further. 
 He believes that the lower temperature aloft is intimately tied to the
existence of radiation,  implying that it is driven by net outwards radiation.
(The effect of solar radiation on the terrestrial atmosphere is not explicitly
mentioned.)  It is difficult to tell whether or not Thomson is in disagreement
with Laplace, but the precision of his statements represents a marked
improvement over his predecessors and his earlier work.   

The principal developers of the field,
Ritter (1878-1883) and Emden (1907), seem to accept the idea of convective
equilibrium. It may be pointed out, however, that this mechanism is in no way
expressed by the equations that these and other authors use to predict the
behaviour of real atmospheres. {\it The concept of convective equilibrium is
introduced to 
 one purpose only: to avoid contradiction with the ideas on thermal equilibrium
of isolated systems. It receives no quantitative theoretical treatment.}

Nor was it accepted by everybody. A famous incidence involves Loschmidt (1876),
who believed that an isolated atmosphere, at equilibrium in a gravitational
field, would have a temperature gradient.  But  arguments presented by Maxwell
and Boltzmann (1896) led Loschmidt to withdraw his objections, which is hardly
surprising given the authority of these two. Nevertheless, it may be pointed out
that no attempt was made, to our knowledge, to settle the question
experimentally. See however (Graeff 2008).

An alternative to convective equilibrium was proposed by Schwarzschild (1906)
and  critically examined by Emden.  To understand how it  works we turn to
Emden's book of 1907, beginning on page 320. Here he invokes a concept that is
conspicuously absent from all his calculations on polytropic spheres in the rest
of the book: heat flow. It must be agreed that the atmosphere is not
completely transparent, and that heat flow is an inevitable consequence of the
existence of a temperature gradient. {\it The most important observation is that
heat flow is possible in stationary configurations ($\dot T = 0$) provided that
the  temperature gradient is constant.} The heat flux due
to conduction and radiation can be expressed as  
$$
\vec F =   C\vec \bigtriangledown T,~~F^i = C^{ij}\p_j T,
$$
where the tensor $C$ includes the thermal conductivity as well as the effective
coefficient of heat transfer by radiation. The divergence of the flux is the
time rate of change of the temperature due to conduction and radiation. In a
stationary, terrestrial atmosphere, with no local energy creation, this must
vanish. Emden's atmospheres are polytropes,
with temperature gradients that are constant (it appears that he takes $C$ to be
constant).
 That is interesting, for it reminds us that the entire edifice
 implicitly demands that this condition, of a constant
temperature gradient,  be satisfied.

We note that the direction of flow is from hot to cold, outwards.  Confining
ourselves to planetary atmospheres, with no local energy generation, this calls
for an explanation, since the ultimate source of energy is above. Here we have
to return to the oldest explanation of the existence of a temperature gradient,
dismissed by Laplace ({\it op. cit.}): the greenhouse effect.  The atmosphere is
highly transparent to the (high frequency) radiation from the Sun  but opaque to
the thermal radiation to which it is converted by the ground. The atmosphere is
thus heated from below! 

If the atmosphere is stable in the sense discussed above, when $\gamma'
<C_P/C_V$, then it is not necessary to assume that any convection takes place. In
this case one speaks of (stable) `radiative equilibrium'. Convective
equilibrium steps in when the stationary atmosphere is unstable, but
it is no longer used to explain the existence of a
temperature gradient; it is the effect rather than the cause of it.  

A difficulty is present in all accounts of stellar structure up to  1920. 
The energy observed to be emitted by the Sun, attributed to 
contraction of the mass and the concomitant release of internal energy, was far
too small to account for the age of the sun as indicated by the geological
record. The situation changed with the discovery of thermonuclear energy
generation. Now there is plenty of energy available. At the same time there
arose the realization that convection sometimes plays a very modest role; the
concept of convective equilibrium was put aside and with it, Kelvin's explanation
of the temperature gradient. According to Eddington (1926), ``convective
equilibrium" must be replaced by ``radiative equilibrium". He does not claim
that this new concept explains the temperature gradient as well as Kelvin's
convective equilibrium does, but in fact  the local
generation of heat by thermonuclear processes creates an outward flow of heat
and a negative temperature gradient.

 Emden's implicit invocation of the heat equation reminds us that this
equation probably should replace the simple rule that `a
 system in equilibrium must have a uniform temperature'.

Finally, there emerges a physical picture that seems to account for all the
principal features of some atmospheres, in a qualitative way. If one begins with
an isolated atmosphere in equilibrium one may take it as an axiom that the
temperature is uniform. A  weak dose of radiation upsets the static equilibrium
and the gas goes into a state of convective equilibrium, characterized by an
adiabatic temperature profile. When the intensity of the radiation increases
it produces an outward heat flux and with it an added pressure. This pressure
is the gradient of the energy of radiation and the additional `potential'
has to be overcome when a parcel of gas is moving downward; that is, the
polytropic index is increased. This in turn stabilizes the atmosphere against
convection and eventually it reaches a state of `radiative equilibrium', with
insignificant convection. See (Cox and Giuli 1968), page 271.  
 If this interpretation is correct then the theory  must be completed by 
inclusion of heat flow as an additional degree of freedom, and by an
account of the properties of the coefficient $C$.  A more distant goal would be
the inclusion of convection into the mathematical desription.

\b\b
\ce{\bf I.7. The mass}

To speak of a definite, isolated physical system we must fix some
attributes, and among such defining properties we shall
include the mass. We insist on this as it shall turn out to be crucial to
the logical coherence of the theory. The density $\rho$ will be taken to
have the interpretation of mass density, and the total mass is the constant
of the motion
$$
M = \int d^3 x ~\rho.
$$
Such integrals, with no limits indicated, are over the domain $\Sigma$ of
definition of $\rho$ and is the total extension of our system in \Rrm$^3$.

Since the total mass is a constant of the motion it is natural to fix
it in advance and to vary the action subject to the constraint
$\int_\Sigma d^3x\, \rho(x) = M$. We introduce a Lagrange multiplier
and the action takes the form
$$
A = \int _\Sigma d^3x\Big(\rho(\dot\Phi - \vec v^2/2) -V\Big)
+\lambda\Big(\int _\Sigma d^3x \rho - M\Big).
\eqno(1.19)
$$
In the simplest case of a polytropic equation of state and no external
forces we get the following equations of motion
$$
\p_\mu J^\mu = 0,~~  V' =  \dot\Phi - \vec v^2/2 +
\lambda   = \hat a\gamma \rho^{1/n}.
$$
Here $\lambda$ is to be chosen for each solution so as to satisfy the
constraint.
In the  case of a static solution with  $\dot \Phi = 0, \vec v = 0$
the density is constant.
    Assuming a finite system with volume ${\cal V}$ we have
$ M = \rho{\cal V} = (\lambda /\hat a\gamma)^n\V$ and since $M$ is given,
$$
\rho =  {M\over{\cal V}} ,~~p = {\hat a\over n}({M\over{\cal
V}})^{\gamma},~~\lambda = \hat a\gamma({M\over{\cal V}})^{1/n}. \eqno(1.20)
$$ 

The conservation of mass has important implications for boundary conditions.
\b\b

 \ve

\no {\steptwo II. The first law}

\ce{\bf II.1. Thermodynamic equilibrium}

A state of thermodynamical equilibrium of a system that consists of a
very large number of identical particles is defined by the values of 3
variables, {\it a priori} independent, the density $D$, the pressure $P$
and the temperature $T$. These are variables taking real values;
they apply to the system as a whole. In the case of any particular system
there is one relation that holds for all equilibrium states, of the form
$$
T = f(D,P).
$$
It is written in this form, rather than $F(T,D,P) = 0$, because a unique
value of $T$ is needed to define a state of equilibrium
between two systems that are in thermal contact with each other: it is
necessary and sufficient that they have the same temperature. This
statement incorporates the zeroth law.

 If we divide our system into subsystems then these will be in
thermal equilibrium with each other only if they  have the same temperature.
This, at least, is the inherited wisdom; we shall honor it as long as
possible.

The ideal gas at equilibrium is defined by global variables $T,D,P$,  and two
relations. The principal one is the gas law
   $$
P/D = {\cal R}T,~~ \R = .8314\times 10^{8} ergs/K,\eqno 
$$
where   $ 1/D$ is the volume of a mole
of gas. The other may take the form of an expression for the internal energy.

  \b\b

\ce{\bf II.2. The ideal gas in statistical mechanics}

Here
again we consider a gas that consists of identical
particles (Boltzmann statistics), each with mass $m$ and subject to
no forces. It is assumed that the $i$th particle has momentum $\vec
p_i$ and kinetic energy $\vec {p_i}^2/2m$.   This is an ideal gas, satisfying
the relation $P/D = {\cal R}T$ at equilibrium. It is assumed that the
number $N$ of particles with energy $E$ is given by the
Maxwell distribution
$$
N(E) \propto {\rm e}^{-E/kT},\eqno(2.1)
$$
which implies a
constant density in configuration space. Now place this gas in a
constant gravitational field, with potential $\phi(x,y,z) = gz,~ g$
constant. Since the potential varies extremely
slowly on the atomic scale it is plausible  that, at
equilibrium,  each horizontal layer ($\phi$ constant) is characterized by
a constant value of the temperature, density and pressure. Since
neighbouring layers are in thermal contact with each other the
temperature must (?) be the same throughout,
$$
T(z) = T = {\rm constant},
$$
and
$$p(z)/\rho(z) = {\cal R}T, ~~
z\geq 0.\eqno(2.2)
$$
The energy of a particle at level $z$ is $\vec  p\, ^2/2m + mgz$ and (2.1)
now implies the following distribution in configuration space,
$$
\rho(x,y,z)  \propto {\rm e}^{-mgz/kT},\eqno(2.3)
$$
   in perfect agreement with (1.17).
   This supports the appearance of the logarithm in the expression for
the potential, Eq.(1.15).  Both derivations of the
distribution rest on the assumption that the temperature is
constant throughout the system.

We conclude that the static solutions of the action principle, with action
density (1.16) and $T$ fixed, describe  the  equilibrium states
of an ideal gas at fixed temperature $T$ in the sense of thermodynamics
and statististical mechanics, even in the presence of the gravitational field,
when no account is taken of radiation. But we do not know
under what conditions the temperature will actually be uniform.

 About this question there has been some debate,
see e.g. Waldram (1985), page 151. It is said that  the kinetic energy of
each atom in a monatomic gas is $3kT/2$ and that, when the temperature is
the same everywhere,  this is paradoxical because it does not take account
of the potential energy of the atom in the gravitational field.    The incident
involving Loschmidt, Maxwell and Boltzmann has already been mentioned. 
\b\b

\ce{\bf II.3. The first law and the internal energy}

Is further generalization possible? Can we extend the model to the case
that the temperature varies with time?
The action must be modified, for the temperature becomes
a dynamical field. Is the  temperature 
one of the variables with respect to which the action must be minimized?
We need an  equation of
motion to predict its evolution. (The heat equation will be discussed later.) The
usual approach is to lay down the additional equation by fiat (Section 1.5); is
this completely satisfactory? Would it perhaps be preferable to have it appear
as the result of minimizing the action with respect to variations of the
temperature field?

To prepare for the generalization we shall examine some of the
main tenets of thermodynamics in the context of the action principle.
The question of whether or not it is profitable to treat the temperature
as a dynamical field variable in the context of the action principle can
best be assessed a little later (Section III.3).

Suppose that the system is in  thermal and mechanical isolation except for
a force that is applied to the boundary. The system is in an equilibrium
state with temperature $T$. The applied force is needed to hold the gas
within the boundary of the domain $\Sigma$, then decreased by a very
small amount leading to a displacement of the boundary and an increase of
the volume by a small amount $d\V$. Assume that this process
is reversible.
The work done by the applied force is
$$
dW = -p d\V.\eqno(2.4)
$$
The first law states that, if the system is in thermal isolation, then this
quantity is the differential of a function $U(T,{\cal V})$ that
is referred to
as the internal energy of the system.

Consider the system that consists of an ideal gas confined to a volume
$\V$ and experiencing no external forces, not even gravitation. If the gas
expands at constant pressure the work done by the gas is $pd{\cal V}$ and
Eq.(1.12) tells us that,
$$
pd\V =   \R T M{d {\cal V}\over {\cal V}}.\eqno(2.5)
$$

The idea of energy conservation suggests a concept of ``internal
energy".
    It is assumed that, under certain
circumstances, the work done by the gas  is at the expence of an internal
energy $U$ so that
$$
pd{\cal V} + dU = 0,
   $$
or
$$
{\cal R}TM d{\cal V}/{\cal V} + dU = 0. %See Saha and others.
$$
It is  an experimental fact (Gay-Lussac 1827, Joule
1850) that the internal energy of an ideal gas is independent
   of the volume (see below) and the more precise statement
   that the internal energy density $u$ is proportional to ${\cal R}T\rho$
is often included in the definition of the ideal gas (Finkelstein 1969,
page 7). Thus
   $$
   u = \hat c_V{\cal R}T\rho,~~ U = \hat c_V {\cal R}TM.
   $$
Statistical  mechanics gives $\hat c_V = n$, where
   $n$ is the adiabatic index and takes the value $
3/2$ for a monatomic gas. Thus
   ${\cal R}TM d{\cal V}/{\cal V} + dU = {\cal R}TM d{\cal V}/{\cal V}
+ n{\cal R}MdT= 0$, which implies that
   $$
   dT = -{1\over n} {T\over {\cal V}} d{\cal V}, ~~T\propto  {\cal
V}^{-1/n}. \eqno(2.6)
   $$
This relates the temperature to the volume and replaces
the statement that $U$ is independent of the volume. The calculation from 
(2.4) onward was done with the understanding that $M = \rho\V$ is fixed.

As we see it,    the
expression for the internal energy in terms of $\V$ and $T$ 
appears to be somewhat {\it ad hoc},  derived from external considerations.
At the deepest level the concept of energy
derives its importance from the fact that it is conserved with the
passage of time, by virtue of the dynamics. The defining
equations of hydrodynamics do not admit a first integral, and no unique
concept of energy; this is a difficulty that our limitation to
laminar flow, and the action principle,
will allow us to overcome. In modern versions of
thermodynamics, and especially in the thermodynamics of irreversible
processes and in radiation thermodynamics, conservation laws are all
important, but they are postulated, one by one, not derived from basic
axioms as is the case in other branches of physics, and they have a purely
formal aspect since they serve only to define various fluxes.
See e.g. Stanyukovich (1960), Castor (2004). 
   \b

  \ve
\ce{\bf II.4. The first law and the hamiltonian}

{\it Having adopted an action principle approach we are bound to
associate the internal energy with the hamiltonian}, but one cannot
escape the fact that the hamiltonian density is defined only up to
the addition of a
constant multiple of the density. When we decide to  adopt a
particular expression to be used as internal energy over a range of
temperatures, we are introducing a new assumption.  Any expression for the
internal energy, together with the implication that applied forces increase
it by an amount determined by the work done, is a  statement about a family
of systems, indexed by the temperature. This cannot come out of the gas law
and implies an independent axiom.

If we adopt the simplest expression for the hamiltonian,
$$
H = \int d^3x (\vec v^2/2 + V), ~~ V = {\cal R}T\rho\log\rho,
$$
to serve as ``internal energy", and repeat the analysis of
   the effect of adiabatically changing the volume by means of an applied 
pressure, then we shall get
$$
pd{\cal V} + dH(T,{\cal V}) = 0, ~~ p = {\cal R}TM/{\cal V}.
$$
In the static case $H = {\cal R}TM\log (M/{\cal V})$ and
$$
dH = {\cal R}M\log(M/ {\cal V}) dT -{\cal R}TMd{\cal V}/{\cal V}.
$$
The second term compensates for $pd{\cal V}$  and so $dT = 0$,
   the temperature does not change.
   This is perfectly consistent with the theory as it has been
   developed so far, but it contradicts experimental results for an 
ideal gas. Besides, variation of our present, isothermal lagrangian with respect
to
   $T$ does not give a reasonable result,
   the lagrangian needs to be improved.

\b\b

\ce{\bf  II.5. The adiabatic  lagrangian}

In this section we shall treat the adiabatic gas; that is, the polytropic
atmosphere with $\gamma' = \gamma$. But at the end we shall make the case for
extending the results to all polytropes.

The two relations $p = \R T\rho$ and $ p = \hat a \rho^\gamma$ imply the
relation $\R T= \hat a\rho ^{1/n}$ between the two independent
variables
$T$ and $\rho$ that holds for a set of configurations related  by
adiabatic transformations.   The  index $n$ may be fixed for all
configurations, while the coefficient $\hat a$ parameterizes the family
of adiabats (polytropes).

It is possible to derive both relations from a principle of
least action, by independent variation of both temperature and density,
but a lagrangian functional of $\Phi,\rho$ and $T$ can only pertain to
a single adiabat. An action principle that describes the whole family of
adiabats must involve additional variables, variables that from a
restricted point of view of the gas appear as sources of heat. {\it A 
lagrangian without sources can be interpreted as applying to a single adiabat
of an isolated system.}

The  application that this study is aimed
  at is the earthly or a stellar atmosphere. Those systems are not isolated,
but they may nevertheless be treated as being formally isolated when the
energy of radiation is included in the hamiltonian and the effect of radiation
is taken into account through the imposition of boundary conditions.  It is
felt that, if external energy sources are going to be invoked, then it is
important that we first establish that there is a need to do so. This is why  we
continue to regard the lagrangian as applying to an isolated system.  

   Two kinds of additions can be made to the lagrangian
   without spoiling the equations of motion that
   are essential to hydrodynamics.

Adding a term independent of $\rho$ and a term linear in $\rho$ we
consider
$$
{\cal L}[\Phi,\rho,T] = \rho(\dot\Phi - \vec v^2/2 -\phi + \lambda )
   -{\cal R}T\rho\log(\rho/ \rho_0)+ \rho\mu[T] + f[T]  .\eqno(2.7)
$$
The additions do not spoil the relation $p = {\cal
R}T\rho$, nor the continuity of the current.   Variation with respect
to $T$ gives
$$
\rho\mu'[T] - {\cal R}\rho\log(\rho/\rho_0) = -f'[T] =
-(4a/3)T^3.\eqno(2.8)
$$
   We have set $f[T] = (a/3)T^4$, in anticipation of the interpretation of
this term as the pressure of the photon gas.   If the
constant
$a$ is the
   Stefan-Boltzmann constant, $a = 7.64\times 10^{-15} ergs/K^4$, as it will
be taken to be, then this term is very small in most circumstances and we
must have, on shell,
   $ 
   \mu'[T] \approx {\cal R}\log(\rho/\rho_0).
   $ 
  The following expression will be used  $$
   \mu[T]  =  n {\cal R} T\log{T\over T_0}.\eqno(2.9) 
   $$
Eq.(2.8) takes the form$$
\R \Big(n+ \log{T^n\rho_0\over T_0^n\rho}\Big) \rho +
{4a\over 3} T^3 = 0, \eqno(2.10)
$$
and in the important case when $n = 3$ 
$$
\R \Big(3+ \log{T^3\rho_0\over T_0^3\rho}\Big){\rho\over T^3} +
{4a\over 3} = 0, 
$$
which is equivalent to Poisson's law $T^3/\rho = $ constant.
This reflects the strong affinity that is found between the
polytropic ideal gas with $n = 3$  and radiation. The value $n = 3$ has
a cosmological significance as well, it is characteristic of the changes
in $\rho, p, T$ induced by uniform expansion (Ritter, Emden, see
Chandrasekhar 1939, page 48). For other values of $n$, Eq.(2.10) is a mild
modification of the polytropic equation of change in the presence of
radiation.

The equation of motion that is obtained by variation with respect to
$\rho$ is
$$
\dot\Phi - \vec v^2/2 -\phi + \lambda +\mu[T]   = {\cal
R}T\big(1+\log(\rho/\rho_0)\big).\eqno(2.11)
$$
Combined with Eq.(2.10) it reduces, in the static case, 
to
$$
 \rho(\phi-\lambda) +
(1+\log k/k_0)\R T\rho  = 0,~~ k := \rho/T^n,   \eqno(2.12)
$$
which has the same form as the 
equation (1.18) studied in Section 1.5. 

 {\it We have thus found an action
that, varied with respect to $\rho,   \Phi$ and $T$, reproduces all the
equations that define the ideal, polytropic gas with $n = 3$. More generally,
for any  value of $n$, it describes a gas that has its effective polytropic
index increased from the `natural' adiabatic value so that the neglect of
convection is justified. In the limit when radiation becomes negligible it fails
to account for the onset of convection, so it should not be used in that case.
But with this exception it is  a mathematical model that incorporates all of the
physical insight that was summarized at the end of Section I.6. 

 We suggest that
using the lagrangian (2.7), with (2.8) and (2.9), is preferable to the usual
assumption that 
$\beta := p_{\rm gas}/p_{\rm tot}$ is constant.} 
Additional evidence for the aptness of the adiabatic lagrangian is found in
the next subsection. 
 
\b\b

\ce{\bf II.6. Energy, pressure and entropy} 

The hamiltonian density is, in the static case, with the choice (2.8)-(2.9),
$$
h = \rho e + \rho\vec v\,^2/2,~~ \rho e  = \phi\rho + \R T\rho\log\big({\rho\over
\rho_0}{T_0^n\over T^n}\big)  -
 {a\over 3}T^4.\eqno(2.13)
$$
  With the aid of Eq.(2.10) we obtain for the hamiltonian, on
shell, when $\phi = 0$ and $\vec v = 0$,
   $$
H_{\rm tot} = n{\cal R} MT +  aT^4\V       ,\eqno(2.14)
$$
   in full agreement
with the familiar expression for the internal energy of an ideal gas with
adiabatic index $n$, augmented by the energy density of the radiation
field. {\it This may be the first time that this
expression for the internal energy has been related to the hamiltonian of an
action principle.}

The pressure was defined alternatively in terms of the potential,
or as the on shell value of the lagrangian.
We prefer to define the total pressure by the requirement that
   $$
   p_{\rm tot}d{\cal V} + dH_{tot} = 0.\eqno(2.15)
   $$
Taking $n = 3$ and $\phi= 0$  we have since $T^3\V$ is constant in this case,
$$
dH_{\rm tot} =
   3RMdT +a(T^3\V)dT = \big[-\R MT/\V - (a/3)T^4)\big]d\V
$$
   and thus
   $$
   p_{\rm tot} = {\cal R}MT/{\cal V} + {a\over 3}T^4.\eqno(2.16)
   $$
This result (2.14-16) is very suggestive.
   It gives the total pressure as the usual pressure of an ideal
gas with polytropic index $n$, augmented by a term that
begs to be interpreted as a pressure due to heat itself, which
is natural when heat is interpreted in terms of electromagnetic
radiation. Its magnitude is one third of the radiative energy
density, as
expected for the photon gas.

   An energy conservation equation follows in standard fashion from the action
principle, namely
$$
{d\over d t}\big(\rho\vec v\,^2/2 +\rho e\big)  + \vec 
 \bigtriangledown \cdot\,(\rho\vec v\,\dot\Phi) = 0.
$$
On the trajectory, $\rho\dot\Phi = \rho\vec v\,^2/2 + h + p$ and the
standard conservation law results (Castor  2004, page 14), in the case that
there is no contribution from heat flow. 

The internal energy density, with the aid of (2.10) can be expressed as
$$
\rho e =    n\R T\rho + {a\over 3}T^4.
$$ 
If $n$ is replaced by $n'$, 
 then this is not in agreement with the energy density for a perfect gas
with adiabatic index $n$:
$$
\rho e =    n\R T\rho + {a\over 3}T^4 + (n'-n)\R T\rho. 
$$ 
  The discreapancy
(the last term) indicates that, when $n'>n$ (the case of convective stability),
if a parcel of air is to  move down in the atmosphere, to higher densities,
then heat has to be supplied at a rate that is proportional to the
change in  temperature. This is precisely the usual interpretation of 
polytropic change. Namely, one considers a change 
$ dT$ in temperature that is accompanied by  an addition of heat $c\rho
dT$. This is a displacement along the polytrope with index 
$$
\gamma' = {C_P+c\over C_V + c},
$$
or $n' = n+c/\R$. Hence
$$
\rho e = n\R T\rho + c\rho T.
$$
The increase in energy is thus accounted for; it implies that there is an extra
supply of energy stored in the gas, distinct from the internal energy and
distinct from the contribution $(a/3)T^4$.  This tallies very well with the
interpretation of the temperature gradient as a greenhouse effect. The extra
energy is supplied by the outgoing flux.

\b\b

\ce{\bf II.7. Virial theorem}

Both (2.14) and (2.16) are usually derived from considerations outside
the proper domain of thermodynamics. We  prefer 
an axiomatic foundation of thermodynamics that is complete in the
sense that it does not need other input. As an  example 
let us discuss the use of the virial theorem to make certain
predictions concerning stability.

The virial theorem was introduced into the present context by Kelvin. It
is based on the scaling properties of the hamiltonian of a system of
particles. If
$H = K+V$, kinetic energy plus potential energy, then the lagrangian is
$K-V$ and the equations of motion imply that, up to a time derivative,
$$
\sum_im_i\dot q_i^2 = 2K = -\sum q_i\p_iV.
$$
In the case examined by Kelvin the potential is homogeneous of degree -1,
so that, \underbar{in the case of periodic motion}, when average is taken
over a period, $V = 2K$.  According to
 Chandrasekhar
 (1938) (pp. 49-51), who also quotes Poincar\'e, the internal energy  
is the kinetic energy associated with  the
microscopic motion of the molecules. It is assumed, usually without
discussion, that the presence of gravitational forces do not affect the  
internal energy, and that the total energy is obtained by simply adding the
gravitational potential energy to it. In the present approach there is no
place for this argument, the hamiltonian is the energy and there is only one
energy.

There is; however, a virial theorem associated with a lagrangian of
the type (2.7), that we abbreviate as
$$
{\cal L} = \rho(\dot \Phi - \vec v^2/2) - \hat V. 
$$
(The potential $\hat V$ includes the gravitational field.) Variation of
$\Phi$ and of $\rho$ give the equations of motion
$$
\dot \Phi = \vec v^2/2+(d\hat V/d\rho),~~ \dot\rho = -{\rm div}(\rho
\vec v),
$$
which implies that
$$
\int d^3x {d\over dt}(\rho\Phi) = \int d^3x\Big(\rho{d\hat V\over d\rho} -
\rho\vec v\,^2/2\Big).
$$
If the system goes through a cycle then the average of this
quantity over the cycle is zero,  
$$
<\int dx~\rho\vec v\,^2/2>~ = ~<\int dx~ \rho{d\hat V\over
d\rho}>.\eqno(2.17)
$$
 In the case of (2.10) we obtain, when $n = 3$,
$$
<\int dx~\rho\vec v\,^2/2>~ = ~<\int dx ~ \Big(\rho(\phi-\lambda) +
4\R T\rho +{4a\over 3}T^4\Big)>.\eqno(2.18)
$$
 With Eq.(2.13) this simplifies to
$$
<\int dx~\rho\vec v\,^2/2>~ = ~<\int dx ~ \Big(\rho(\phi-\lambda) +
\R(1+\log k)\rho T\Big)>.\eqno(2.19)
$$

 {\it This result, like classical virial theorems, applies
exclusively to the case of periodic motion.}

In the special case $\vec v = 0$ Eq.(2.19) is a direct consequence of the
equations of motion. Such relations, that do not depend on the periodicity
of the motion, are not true virial theorems.

\ve
 \ce {\bf II.8. The centrifuge and the atmosphere}

Kelvin justified the polytropic model of the atmosphere in terms of radiation
and convection. Eddington discounted the role of convection and relied on a
concept of radiative equilibrium.  To find out what happens \underbar{in the
case of complete insulation} we study the analogous situation in a
centerfuge. 

Consider an ideal gas. By a series of experiments in
which gravity does not play a role,  involving reversible
changes in temperature and pressure, it is found that, at equilibrium, the
laws $p/\rho = \R T$ and $\rho = kT^n$ are satisfied, constants $k,n$
fixed. When supplemented by the laws of hydrodynamics, they
are found to hold, or at least they are strongly believed to hold, in
configurations involving flow, over a limited time span, in the absence of
external forces. 
In addition it is said that, at equilibrium, the temperature must
be  unifoem. Keeping an open mind,
  let us refer to this last statement as ``the axiom".   
We are talking about a fixed quantity of gas contained in a vessel,
the walls of which present no friction and pass no heat.

Let the walls of the vessel be two vertical, concentric cylinders,
and  construct a stationary solution of the equations of motion. And why not?
We have experimental confirmation of the equations of motion, we applied them to
the theory of sound with a degree of confidence that is so high that  
 the   prediction of rapid variations in temperature may never have been
subjected to verification (?).   In terms of cylindrical coordinates, take
$v_z = v_r = 0, v_\theta =
\omega$, constant. The continuity
equation is satisfied with $\rho$ any function of $r$ alone. Then neither
$T$  nor
$p$  is constant, for the hydrodynamical equations demand  that
 $$
r\omega^2 = cT', ~~ c = (n+1)\R \approx  10^7 cm/sec^2 K~~
{\rm (for~ air)}.
$$ 
At first sight, this seems to violate the axiom, but perhaps not,
 for this is not a static configuration. 
To save the axiom let us suppose that, by conduction, convection or
radiation,
 the temperature will tend towards uniformity. Perhaps after a suitably long
time has passed, $T$ has become constant, in violation of
the equations of motion.   Let us remember that no
heat or any other influence is supposed to go by the walls; then surely
energy and angular momentum must both be preserved during the time that the
temperature is leveling out. It also seems reasonable to assume that the
final configuration is (macroscopically) stationary and uniform, since the
existence of fluctuations would imply that the entropy had not reached its
maximum. But  a stationary state with non zero density gradient and uniform
temperature would seem to contradict the assumptions that we made about the gas,
which makes the existence of such a state somewhat problematic.

If we also accept the equivalence principle,
then from the point of view of a local observer 
at rest in the flow there is a centrifugal force field, a density gradient
and, by the laws of Poisson, 
a temperature gradient.  The equivalence principle only applies to conditions
at one point, and one can question whether the gradient of the temperature
or of the density  is sufficiently local to be covered by the principle. The
entire theory of relativistic thermodynamics has been founded on the
belief that it is (Tolman 1934). 

If we do accept the equivalence principle 
 (without necessarily embracing the tenets of  traditional
relativistic thermodynamics), then we shall be lead to expect that a vertical
column of an ideal gas, in mechanical equilibrium under the influence of
terrestrial gravity, \underbar{and perfectly isolated},  will have a
pressure and temperature gradient exactly of the form predicted by Lane. 
This seems to contradict what we think is the prevailing opinion of
atmospheric scientists, that the temperature gradient owes its existence to
the heating associated with solar radiation.  

Further measurements in the atmosphere are unlikely to throw light on this,
since isolation is out of the question. Experiments with a centrifuge
may be more realistic.  The temperature lapse rate is $r\omega^2\times
10^{-7} K/cm$. If the acceleration is 1000 $g$ at the outer wall,
then the lapse rate will be .1$K/cm$. The temperature difference between
the inner and outer walls will thus be  1 $K$ if the distance is 10$cm$.
In a practical experiment one does not have the gas flow between
concentric, stationary cylinders. Instead a tube filled with the gas is
oriented radially on a turntable. Friction against the walls is thus
eliminated and heat loss is much easier to control.  

On purely theoretical grounds we have come to doubt that complete
equilibrium implies a uniform temperature in all cases. In fact, Tolman (1934,
page 314) shows that, according to General Relativity, the  temperature of an
isolated photon gas in a gravitational field is not quite uniform. The predicted
magnitude of this effect is very small, but it shows that there are
circumstances in which statistical mechanics is not the absolute truth.

The strongest argument that we have found, against the indefinite persistence of
a temperature gradient in an isolated system, is this. Imagine a large heat bath
located in the region $z>0$. A vertical tube, filled with an ideal gas, has its
upper end in thermal contact with the bath, otherwise it is isolated. Assume
that, at equilibrium, the lower part of the tube has a temperature that is
higher than that of the bath, then we can run an engine by taking out a small
amount of heat from the bottom of the tube and returning it to the bath. In
order for equilibrium to be restored, the heat thus taken out of the tube will
have to be be restored from the bath, which implies a  spontaneous transfer of
heat in the directio opposed to the gradient of the temperature, in
violation of one of the statements of the second law (Clausius 1887). According
to Graeff (2008), this is exactly what he observes!

If our analysis of the
centrifuge is incorrect, what is the mistake? The conventional view is that, if
a temperature gradient exists at one time, then heat flow will tend to dissipate
it. Dissipation is determined by Fourier's heat equation; in its general form
$$
{D\over Dt} T = \p_iC^{ij}\p_j T.\eqno(2.20)
$$
Our error consists of the fact that we are using a lagrangian that
contains neither space derivatives nor time derivatives of the temperature.
We are neglecting phenomena that have a long time scale, such as conduction
and dissipation .  This can be justified if we limit our attention to stationary
configurations, when $\dot T = 0$, if the temperature distribution makes the
right hand side of the equation equal is zero. But a more complete understanding
of the phenomena requires that we consider nonstationary situations as well, and
for that we shall have to include additional terms in the action, terms that
become significant when radiation is not.

It is not just the paradox of the centrifuge that forces us in this direction.
It is important that the theory should account for a situation in which the
radiation is subject to temporal variations, as  is certainly the case  
for the earthly atmosphere.

\b\ve
\ve
\no{\steptwo III. Sources}

\ce{\bf III.1. Generic source}

We have proposed to extremize the lagrangian with respect to all three
fields, $\Phi, \rho$ and $T$. We have found an expression for the
adiabatic lagrangian density, the Euler-Lagrange equations of which pass the
two tests: 1. When the effect of radiation pressure is neglected they give
precisely the equations that govern the polytropic atmosphere, for all $n$.   2.
In the case that $n = 3$ and with the radiation term included, they reduce to the
equations of radiative equilibrium.  
 
 The adiabatic lagrangian describes a single adiabat.  
To remove this limitation in a formal way, let us add another
 term to the lagrangian density,
$$
{\cal L} = \rho(\dot\Phi - \vec v\,^2/2 -gz + \lambda )
 -{\cal R}T\rho\log {k\over k_0}+   f[T] +  \rho TS,
 \eqno(3.1)
$$
where $S$  is an external source. The factor $\rho$ in the source term
 is natural and the factor $T$ is chosen to make $S$ play the role of a
local adiabatic parameter.  We have introduced the variable $k$  and the
parameter $k_0$ by
$$
\rho = kT^n,~~ \rho_0 = k_0 T_0^n,~~(k_0 \rightarrow 1);
$$
Then $k^{-1/n}$ is Emden's ``polytropic temperature".
It will be recalled that $k_0$ parameterizes a family of adiabats; in
fact, for an isothermal expansion, the variation of $-\R \log k $ is
precisely the change in specific entropy. The introduction of the source
$S$ turns $-\R\log k_0$ into a field with the interpretation of entropy.
    We no longer need the parameter
and so, following Lane, we shall use units of density such that
$k_0 = 1$.

The internal specific entropy is $\R\log (T^n/\rho)$ and the total specific
entropy is
$$
S_{tot} =  \R\log {T^n\over\rho} + S.
$$
With this convention
$$
{\cal L} = \rho(\dot\Phi - \vec v^2/2 -\phi +\lambda)
 +    \rho  TS_{tot} + {a\over 3}T^4,\eqno(3.1)
$$
where $\phi$ is the gravitational potential. 
Variation with respect to $T$ leads to
$$
\rho{\p\over \p T} (TS_{tot}) +{4a\over 3}T^3 = 0.\eqno(3.2)
$$
As an equation for $S_{tot}$ it has the general solution
$$
S_{tot} = -{a\over 3\rho}T^3 - {1\over \rho T} V[\rho].\eqno(3.3)
$$
Taking this as the definition of the potential ($V$ is the value of
$\rho  TS_{tot} + {a\over 3}T^4$ at the extremum with respect to variation of
$T$) we have
$$
\L = \rho(\dot\Phi - \vec v\,^2/2 -\phi + \lambda)  -V[\rho].\eqno(3.4)
$$
The gradient of the equation obtained by variation of $\rho$ is
$$
-\rho{D\over Dt}\vec v - \rho\,{\rm grad}\,\phi = {\rm grad}\, p,\eqno(3.5)
$$
with
$$
 p =\rho V' - V = (1-\rho {d\over d\rho})(\rho TS_{tot} + {a\over 3} T^4)
 = -T\rho^2{\p S_{tot}\over \p \rho}+ {a\over 3}T^4 .
 \eqno(3.6)
$$ 
The last equation is justified by the fact that the partial
 derivative of $\rho TS_{tot} + (a/3) T^4$ with repect to $T$ vanishes, Eq.(3.2).

We shall verify some important relations of thermodynamics,
 and for this we must take $T$ and $\rho$ to be constant, with $M = \rho\V$,
and $\phi = 0$.  In this case
$$
p = MT{\p S_{tot}\over \p \V}+ {a\over 3}T^4 =  \R M T/\V + {a\over 3}T^4
 + MT{\p S\over \p \V}. \eqno(3.7)
$$
The hamiltonian density is, in the static case, in the absence of gravity,
$$
h =   - \rho T S_{tot} - {a\over 3}T^4,~~{\rm implying ~that}~~ U
 = -MTS_{tot} - {a\over 3}T^4\V.\eqno(3.8)
$$
Variation of $h$ with respect to $T$ gives zero on shell, so this
  is the same as
$$
u = (1-T\p_T)h =  \rho T^2 {\p S_{tot} \over \p T}  + aT^4
 = n\R\rho T + aT^4 + \rho T^2{\p S\over \p T}.
$$
Thus 
$$
U = MT^2 {\p  S_{tot} \over \p T}  + aT^4\V
 = n\R M T + aT^4\V + MT{\p S\over \p T}.\eqno(3.9)
$$
Using (3.7) and (3.8) one verifies that
$$
{\p U\over \p\V}  = (T\p_T - 1)p,\eqno(3.10)
$$
 an important  consequence  of the existence of entropy in general. See
 Finkelstein (1969)  page 26. 
Also,
$$
dU = MT {\p  S_{tot} \over \p T}  dT + Td\Big( MT{\p  S_{tot} \over \p T} 
 \Big) + 4aT^3\V dT + aT^4d\V,
$$
$$
pd\V = MT{\p S_{tot}\over \p\V}d\V + {a\over 3}T^4d\V,
$$
and the sum is
$
dU + pd\V = dQ = T(\p p /\p T) = 0$,  the last on shell.

We used the last expressions in (3.7) and (3.9) because they are familiar,
 but if we return  to (3.8) and the first expression for $p$ in (3.7) we see
immediately that $dU + pd\V = 0$. 

If instead we consider a change that involves outside forces acting via
 the source, then
$dU + pd\V = MT\delta S$, which confirms the interpretation of $S$ as a
contribution to the specific entropy.

The calculations that have been presented in this subsection are
offered as proof that the variational approach that is being advocated  
is fully compatible with classical thermodynamics. This gives us
faith in the basic framework and courage to proceed on a more speculative
course.  
\b\b

\ce{\bf III.2. Electromagnetic fields}  

We write the Maxwell lagrangian as follows,
$$
{\cal L}_{\rm rad} = {1\over 2\epsilon} \vec D^2 - {\mu\over 2}\vec H^2 +
 \vec D\cdot (\vec \p A_0 - \dot{\vec A}) -\vec H\cdot \vec \p\wedge\vec
A + JA,\eqno(3.13)
$$
and add it to the ideal gas lagrangian
$$
{\cal L}_{\rm gas} = \rho(\dot\Phi - \vec v^2/2 -\phi + \lambda )
 -{\cal R}T\rho\log k +    {a\over 3}T^4,
 \eqno(3.14)
$$
 Since  the susceptibility of
an ideal gas is small, the dielectric constant may be expressed by
$$
\epsilon = 1 +\kappa[\rho,T] , ~~{\rm or}~~ {1\over \epsilon} =
 1 - \kappa[\rho,T].\eqno(3.15)
$$
Paramagnetic effects will be ignored at present.
An interaction between the two systems occurs through the dependence of the
susceptibility on $\rho$. The source $S$ has become $-(\vec
D^2/2\rho)(\kappa/T)$. If this quantity has a constant value then it produces a
shift in the value of the  parameter $k$.

 Two interpretations are possible. The electromagnetic field may represent
an external field, produced mainly by the source $J$, and affecting the gas by
way of the coupling implied by the dependence of the dielectric constant on
$\rho$. Alternatively, $J=0$ and the field is produced by microscopic 
fluctuations, quantum vacuum fluctuations as well as effects of the
intrinsic dipoles of the molecules of the gas. In this latter case the main
effect of radiation is represented by the radiation term $aT^4/3$.  Our
difficulty is that neither interpretation is complete, and that we do not
have a sufficient grasp of the general case when either interpretation is
only half right. The following should therefore be regarded as tentative. 
 
Variation of the total action, with lagrangian ${\cal L}_{\rm rad} +
 {\cal L}_{\rm gas}$, with respect to $\vec A, \vec D, \vec H$ and $T$ gives
$$
\dot{\vec D} = \vec \p\wedge \vec H, \eqno(3.14)
$$
$$ 
\dot{\vec A} = \vec D/\epsilon,\eqno(3.15)
$$
$$
\mu\vec H = -\vec\p\wedge \vec A,\eqno(3.16)
$$
and
$$
\R(n-\log k)\rho -{\vec D\,^2\over 2}{\p\kappa\over \p T}+
 {4a\over 3}T^3=0.\eqno(3.17)
$$
Taking into account the first 3 equations we find for the static hamiltonian
$$
H =  \int d^3 x \Big( \phi\rho + \R \rho T \log k + {\vec D^2\over
2 } +{\mu \vec H^2\over 2}  -{\vec D^2\over 2}{ \kappa\over  T}-
{a\over 3}T^4 \Big).
$$ 
 With the help of (3.17) it becomes
 $$
 H = \int d^3 x\Big( \phi\rho + n\R \rho T + {\vec D^2\over 2 } +
 {\mu \vec H^2\over 2} + a T^4\Big) -\int d^3 x \,T{\vec D^2\over 2}{\p (T 
\kappa)\over \p T}.\eqno(3.18)
 $$
 The last term, from the point of view of  the thermodynamical
 interpretation of electrostatics, is 
 recognized as the entropy (Panofsky and Phillips 1955). On  a suitable 
choice of the functional
 $\kappa$ it merges into the 
 internal energy. For example, if $\kappa = \rho T$ it takes the form
 $\rho T S$ with $S = \vec D^2 $.
\b\b

\ce{\bf III.3. Discussion 1. Using  $T$ as a dynamical variable}

The  idea of extremizing thermodynamical potentials with respect to the
temperature is far from new, but  in the context of 
the action principle it is likely to raise questions. 

 Radiation Hydrodynamics (Castor 2004) 
 is defined by the following equations (without specialization
to laminar flow): the continuity equation, the hydrodynamical equation, the
equation of motion, and an `energy equation'. This last
equation takes the place of the equation that results from
variation of the temperature. To see this we have only to review the
canonical conservation of energy in lagrangian/hamiltonian form.

With  ${\cal L}$ as in (3.4),
$$
 {\cal L}  = \rho(\dot\Phi - \vec v\,^2/2 - \phi + \lambda) - V[\rho,T], 
$$
with no derivatives in the functional $V[\rho,T]$. We have
$$
{d {\cal L}\over dt}  = \dot\rho {\p{\cal L} \over \p\rho} + \dot\Phi{\p{\cal
L} \over \p\dot\Phi} + {\rm grad}\,\Phi\cdot {\p{\cal L} \over \p\,{\rm
grad}\,\Phi}  + \dot T{\p{\cal L} \over \p T}.
$$
On shell, the first term on the right is zero. The equation can be
rearranged to read
$$
{D h\over D t} + {\rm div} (p \,\vec v) = - \dot T {\p  {\cal L}\over \p
T}.
$$
The `energy equation' is the vanishing of the left side; it is thus
equivalent to setting $\p  {\cal L}/\p T = 0$. There is no evidence
of a contribution to energy transport by heat flow, but that is not a problem if
we accept Emden's postulate of a constant heat flow.

The effect of heat transfer by radiation and by conduction has been left out,
for simplicity and since it does not immediately affect the polytropic
atmosphere. But it means that our treatment is incomplete. Most important, it 
prevents us from dealing with the isothermal atmosphere that is, after all,
an important special case. Consequently, it is felt that a general
understanding of the effect of radiation is lacking.

Let us examine the total lagrangian, 
$$
{\cal L} = {\cal L}_{\rm rad} + {\cal L}_{\rm gas} =
 \rho(\dot\Phi - \vec v\,^2/2 - \phi + \lambda) - \R T\rho\log{\rho\over
T^3} 
$$
$$
 + {\vec D^2\over 2\epsilon} + {\mu\over 2}\vec H^2 +
 \vec D\cdot (\vec \p A_0 - \dot{\vec A}) -\vec H\cdot \vec \p\wedge\vec A+
JA +{a\over 3}T^4.\eqno(3.19)
$$
So long as $\epsilon, \mu$ and $J$ are independent of $\rho,T$ and $\vec
v$, the variational equations of motion that are obtained by variation of
 $\vec v, \rho, \vec A, \vec H$ and $\vec D$ are all conventional, at least
when $n = 3$ (for  all $n$ if radiation is neglegible). It would be possible
to be content with that and fix
$T$ by fiat, as is usual;  in the case of the ideal gas without radiation the
result is the same.  But if $\epsilon$ depends on $\rho$ and  on $T$,
which is actually the case, then we get into a situation that provides the
strongest justification yet  for preferring an action principle formulation
with
$T$ as a dynamical variable. The equations of motion include a contribution
from the variation of $\epsilon$ with respect to $\rho$, so that one of the
basic hydrodynamical equations is modified. Thus it is clear that the
extension of the theory, to include the effect of radiation, is not just a
matter of including additional equations for the new degrees of freedom. The
presence of the term $\vec D^2/2\epsilon[\rho,T]$ certainly introduces the
density
$\rho$ into Maxwell's equations; that it introduces $\vec D$ into the
hydrodynamical equations is clear as well. {\it The over all
consistency of the total system of equations can be ensured by heeding
 Onsager's principle of balance, but the action principle makes it
automatic.}

Variation of the action with respect to $T$ offers additional advantages.
The usual procedure, that amounts to fixing $ \rho = kT^n$, $k$ and $n$
constant, gives the same result when radiation is a relatively unimportant
companion to the ideal gas, but in the other limiting case, when the density
is very dilute and the gas becomes an insignificant addition to the photon
gas, it is no longer tenable. We need an interpolation between the two
extreme cases and this is provided naturally by the postulate that the
action is stationary with respect to variations of the temperature field.

  In the absence of the ideal gas
we have another interesting system, the pure photon gas. The analogy between
the photon gas and the ideal gas is often stressed; there is an analogue of
the polytropic relation that fixes the temperature in terms of
$\rho$; the pressure of the photon field
 is   $(a/3) T^4$. Our lagrangian already contains this pressure; we should
like to  discover a closer connection between it and the electromagnetic
field.   In the limit when the density of the ideal gas is zero, Eq.(3.17)
becomes
$$
 -{\vec D\,^2\over 2}{\p\kappa\over \p T}+ {4a\over 3}T^3=0.  
$$
In the absence of the gas it is reasonable to impose Lorentz invariance, so
we include magnetic effects by completing the last to
$$
 -{ F\,^2\over 2}{\p\kappa\over \p T}+ {4a\over 3}T^3=0.  
$$
If we suppose that
$\kappa[\rho,T]$, in the limit
$\rho = 0$,
 takes the form  $\alpha T^2$, then
$$
 \alpha    F^2 = {4a\over 3}T^2.
$$
The radiation from a gas of Hertzian dipoles can be shown, with the help of
the Stefan-Boltzman law and Wien's displacement law, to satisfy a relation of
precisely this form. Whether the same relation holds in vacuum is uncertain,
but it is suggested by an analysis of the effective Born-Infeld lagrangian
calculated on the basis of the scattering of light by light (Euler 1936, 
Karplus and Neuman 1950). See also McKenna and Platzman (1962).
 \b\b

\ce{\bf III.4. Discussion 2. The temperature gradient of the atmosphere}

We return to the question of the heating of the atmosphere by solar
readiation.  The susceptibility of air may be approximated by $\kappa =
\alpha\rho /T$ ($\alpha$ constant); in which case the lagrangian includes the
following
$$
-\R T \rho\log\rho + n\R \rho T\log T + \alpha\rho T^{-1}\vec
D^2/2.\eqno(3.20)
$$
Up to a factor  $\rho T$, the first term is  the entropy of the
isothermal gas. The first two terms together give the entropy of the
polytropic gas. The entropy introduced by radiation is $ \alpha  \vec
D^2(\p(\kappa T)/\p T)$, namely zero under the assumptions made, 
which would suggest that the greater effect of radiation on the specific
entropy of air comes about through the influence of the term $aT^4/3$ in the
lagrangian. The second term was included in our theory in order that
variation of the action give both of the Poisson relations, the ideal gas
law and $\rho \propto T^n$.  Let us return to the problem of determining the
origin of  the temperature gradient, the physical reason for the fact that the 
atmosphere is nearly adiabatic. In the model, adiabacity comes with the
inclusion of the term $n\R \rho T\log T$ in the action. If this term could be
linked with the last term in (3.20); that is, if it could  be shown that  
$\alpha \vec D^2/2 \approx n\R T^2\log T$, then we could conclude that the
temperature profile is created by the radiation. But this seems
unlikely, as it amounts to fine tuning.  If we suppose that the term in
question is responsible for  the bulk of the effect of solar radiation on the
atmosphere of the earth, then we must admit that it undergoes important diurnal
and seasonal variation. In fact, the polytropic index of the troposphere is
usually quoted as a constant, without any indication that important variations
have been observed over a period of time.

It could be argued that the relatively stable polytropic index
of the earth is evidence that the atmosphere, if
isolated, would continue to manifest a temperature gradient and that an ideal
gas, isolated in a gravitational field, may not tend to an equilibrium state
with uniform temperature, but the explanation of the temperature gradient in
terms of convection is more attractive by far. See the last paragraph of
Section I.6. Additional terms (containing in particular derivatives of the
temperature) are needed in the lagrangian, although they are insignificant in
applications to the polytropic atmosphere under conditions of strong radiation
fields.

\b

\ve 

\no{\steptwo  IV. Stability of atmospheres}

\ce{\bf IV.1. The isothermal  column}

We consider the space that is tangent to a static solution with density
$\rho_0$.  Setting 
 $
\rho = \rho_0 + \delta\rho
 $
we have the following equations for the perturbation $\delta\rho$,
$$
-\dot v = \R T(\delta\rho/\rho_0)',~~ \delta\dot\rho = -(\rho_0 v)',
$$
where the prime denotes differentiation with respect to $z$. Thus
$$
\delta\ddot\rho = \R T(\rho_0\alpha')',~~ \alpha :=
\delta\rho/\rho_0.\eqno(4.1)
$$
For a harmonic mode with frequency $\omega$,
$
-\omega^2\delta\rho =  R T (\rho_0\alpha')',$ and
$$
-{\omega^2\over \R T}\int \alpha^2\rho_0dz = \int\alpha(\rho_0\alpha')'dz
 = \delta\rho\,\alpha'\Big|_0^\infty - \int \rho_o \alpha'^2 dz.
$$
The configuration is stable if this implies that $\omega^2>0$, which will
 be the case if the boundary term vanishes. To justify any choice of
boundary conditions we have only the conservation of mass, 
$\int\delta \rho dz = 0$. This ensures that $\delta\rho$ fall off at
 infinity and we are left with $-\delta\rho(0)
\alpha'(0)$.

We shall show that $\alpha'(0) = 0$. Eq.(4.1) tells us that 
$$
-{\omega^2\over \R T}\alpha =    (\rho'_0/\rho_0)\alpha' + \alpha'' 
= -{g\over \R T} \alpha' + \alpha''.
$$
This is a linear differential equation with constant coefficients, 
with general solution
$$
\alpha =  A\e^{k_+z} + B\e^{k_-z},~~ k_\pm =  
{g\over 2\R T} \pm \sqrt{({g\over 2\R T})^2 - {\omega^2\over \R T}}.
$$
Since, up to an irrelevant constant factor, $\rho_0 = \exp(-gz/\R T)$,
$$
\delta\rho = \rho_0\alpha =  A\e^{a_+z} + B\e^{a_-z},~~ a_\pm =  - {g\over
2\R T}\pm \sqrt{({g\over 2\R T})^2 - {\omega^2\over \R T}}. 
$$
These functions are integrable only if $0<\omega^2 <g^2/4\R T$, and in that
 case 
$$
\delta M = \int\delta\rho dz = -{A\over a_+} - {B\over a_-} = {A\over k_-} +
{B\over k_+},
$$
the vanishing of which requires that $\alpha'(0) = 0$. When $\omega^2 >g^2/4\R T$ we have instead
to do with a contour integral, and reach the same conclusion. Therefore, not
only is the condition $\omega^2 >0$ verified; it is  also confirmed that the
boundary condition $\alpha'(0) = 0$ is the only one possible. We have seen
that this choice of boundary conditions is the one that ensures the
conservation of mass.

\b\b

\ce{\bf IV.2. The polytropic column}

Let us leave the parameter $k =\rho/T^n$ free and fix the value of $n$. This
conforms to the usual approach when the temperature is fixed by edict, but
it is consistent with our formulation if $n = 3$ only. We study the
stability to vertical perturbations.

The static solution is
$$
cT = \lambda-gz,~~ c := \R(1+\log k).
$$
A first order perturbation satisfies
$$
\dot\Phi + \delta\lambda = c\delta T,~~{\rm thus}~~\dot v =- c\,\delta
T',\eqno(IV.1)
$$ 
and
$$
\dot\rho = -({v\,\rho})',~~ \ddot \rho = -(\dot v \rho)'
,~~{\rm and}~~
nT^{n-1}d T   = c(T^n\delta T')'.\eqno(IV.2)
$$ 
Let $x = \lambda/g - z,~ 0<x<\lambda/g$ and let $f' = df/dx$ from now on.
Solutions of the type $\delta T = \exp(i\omega t)f(x)$
satisfy the equation
$$
(x^n\delta T')' + {\nu^2\over x}(x^n\delta T) = 0, ~~ \nu^2 = 
n\omega^2/g.\eqno(IV.3)
$$
The solution that is regular at the origin of $x$ (the top of the
atmosphere) is  
$$
\delta T = \,  _0\hskip-.5mmF_2(n,-\nu^2x)\e^{i\omega t}.
$$
   The generalized hypergeometric function is
positive for positive argument and it oscillates around zero for negative
argument.
\b
\no\underbar{Boundary conditions}. If we fix $\delta T = 0$ at the bottom of
the column we can prove stability as follows. For a harmonic perturbation,
$$
 \nu^2\int x^{n-1}(\delta T)^2 dx = -\int (x^n\delta T')'\delta T dx
=
\int x^n(\delta T')^2dx > 0,
$$
which shows that $\nu^2$ is positive and that the solutions are oscillatory
in time. But there is no justification for this choice of boundary condition.

It is not unusual to fix the upper boundary, and to require that the
perturbation vanish there. If
$\nu^2$ is positive the argument of the hypergeometric function is negative.
The function oscillates around zero and for a discrete set of values of
the frequency it vanishes at the upper end. If
$\nu^2$ is negative then the hypergeometric function is positive, and the
boundary condition cannot be met.

However, there seems to be no better reason to fix the upper boundary. The
natural boundary condition is that the mass must be preserved, thus
$$
\delta M = \int\delta\rho dx = \int T^{n-1}\delta Tdx = 0.
$$
This may happen for a discrete set of positive values of $\nu^2$. 
For negative values of $\nu^2$ the integrand is definite so that it can not
happen. The calculation is valid only in the case $n = 3$; this atmosphere is
stable. For other values of $n$ the calculation is more difficult.

  The problem can be
converted to a standard boundary value problem by rescaling of the
coordinate.  
 
The mass is
$$
M = Ak({g\over a})^3\int d x\,x^3 = {Ak\over 4} ({\lambda\over
a})^{3}{\lambda\over g}.\eqno(IV.4)
$$
The proper definition of gravitational energy is ambiguous, but 
 (II.21) suggests that it is 
$$
E_g =  Ag\int \rho(gz-\lambda )\,dz = -  Akg({g\over a})^3 
\int_0^{\lambda/g} dx\, x^4 = -{Ak\lambda \over
 5}({\lambda\over a})^{3} {\lambda\over g} 
$$
The last expression, and those that follow, refer to the static solutions.
The thermodynamic part of the hamiltonian is
$$
H - E_g = A\int r^2dr ( \R T\rho\log k + {a\over 3}T^4) = 
{  Akc\over 4} \int_R^{ \lambda/g}dx\, T^4
$$
 $$
= { Ak\lambda \over 20} ({\lambda\over a})^3 {\lambda\over g} ={-1\over 4}
E_g.\eqno(IV.5)
$$
 The integrand on the right hand side of (II.21) is thus $E_g + 4(H-E_g) =
0$, as it must be.  

\b\b

\ce{\bf IV.3. The polytropic gas sphere. The hamiltonian}

Here we study the self gravitating polytropic gas. A correction is needed
 in the expression for the lagrangian, and we need to take care with respect
to the definition of the gravitational potential.

First of all, it would not be difficult to argue that the correct expression
 for the gravitational energy is
$$
E_g = -{G\over 2} \int d^3xd^3x'
{\rho(\vec x)\rho(\vec x\,')\over |\vec x - \vec x'|},\eqno(IV.6)
$$ 
and that, consequently, the term $-gz$ in the lagrangian has to be
replaced by $-E_g$.  Nothing else is needed, but to make an important point
it will be useful to introduce a potential, a functional $\phi[\rho]$
defined by
$$
\phi[\rho](\vec x) = \phi(\vec x) = \int d^3x 
{G\rho(\vec x\,')\over |\vec x- \vec x\,'|} + \phi[0].\eqno(IV.7)
$$
The last term, $\phi[0] := \phi[\rho]|_{\rho = 0}$ is of course an arbitrary
constant field. The value chosen for this constant is irrelevant, but it
must be kept in mind that it is chosen once and for all and that \underbar{it
is independent of $\rho$}. The sign is opposite to that used by Eddington; it
is chosen so that the gravitational force is -grad $\phi $. 

For any spherically symmetric distribution define
$$
\M(r) =  4\pi\int_0^r  r'^2\rho(r');\eqno(IV.8)
$$
then
$$
\phi(\vec x)\rightarrow \phi(r) = \int_0^r {G\M\over r'^2} dr' + \phi(0),
\eqno(IV.9)
$$
$$
r^2\phi' = G\M,~~ \M' = 4\pi r^2\rho.\eqno(IV.10)
$$
Since first offered by Perry (1899) it has become customary  to
present a criterion 
for the stability of static configurations, based on an
evaluation of the energy. The better to understand it we replace the 
definition (4.6) of the gravitational energy by
$$
E_g = {1\over 2}\int \rho\phi d^3x = {1\over 2}\int\M'\phi \, dr,\eqno(IV.11)
$$
acknowledging that, since $\phi$ is defined up to an additive constant,
 the same is true of the energy.

As we have seen in Section II.7, it is reasonable to identify the
total energy (including the gravitational energy and the internal energy) 
with the hamiltonian. The lagrangian is  
$$
L = \rho(\dot\Phi - \vec v\,^2/2 - \phi/2 + \lambda) - \R T \rho \log k
 - {a\over 3}T^4.\eqno(IV.12)
$$
The factors of 1/2 in the term $\rho\phi/2$  and in (IV.11) arise from the
fact that $\rho\phi$ is a homogeneous functional of $\rho$ of order
2, $\delta(\rho\phi/2) = \phi$. Variation of the action with respect to  $T$
gives the relation
$$
\R T(n-\log k) = {4a\over 3} {T^4\over \rho}.\eqno(IV.13)
$$
The hamiltonian density is
$$
h = \rho\vec v\,^2/2 + {1\over 2}\rho\phi + \R T\rho\log k + {a\over 3}T^4,
$$
or, in view of (IV.13),  
$$
h ={1\over 2}\vec v\,^2 + {1\over 2}\rho\phi + {3c\over 4}T\rho,~~ c =
\R({n\over 3} + \log k),
$$
whence the hamiltonian (= total energy)
$$
H = \int d^3x h = E_g + \int d^3x({1\over 2}\vec v\,^2 + {3c\over
4}T\rho).\eqno(IV.14)
$$
In the static case we have the equation of motion
$$
\lambda = \phi + cT.\eqno(IV.15)
$$
Since the surface of the star is at the point where $T = 0$, it follows that
$$
\lambda = \phi(R) = \phi(0) + cT(0).\eqno(IV.16)
$$
Applying $\int d^3x\rho$ to (IV.15) we obtain
$$
M\lambda = 2E_g + \int cT\rho d^3x.
$$
Hence (4.14) reduces, in the static case, to
$$
H = E_g + {3\over 4}\big(M\phi(R) - 2E_g\big) = -E_g/2 + {3\over 4}M\phi(R).
\eqno(IV.17) 
$$
Note that $H = E_g + U$ and that this is  what Eddington and others
 call the total energy.

It remains to calculate $E_g$, and here we follow Eddington. To begin, using
 only the definitions (4.8)- (4.11), 
$$
E_g = 4\pi\int_o^R{1\over 2}\rho\phi r^2dr = {1\over 2}\int\M'\phi dr = 
{1\over 2}M\phi(R) - {1\over 2}\int\M\phi'dr,\eqno(4.18)
$$
and
$$
-{1\over 2}\int \M \phi' dx = -{1\over 2}\int{G \M^2\over r^2}dr = 
{1\over 2}\int G \M^2({1\over r})'dr = {1\over 2}{GM^2\over R} - \int{G\over
r}\M d\M.\eqno(4.19)
$$
Next, the polytropic relation can be used to show that  
$$
{3\over 4}\int\M\phi'dr = \int{G\over r}\M d\M.
$$
Hence $\int\M\phi' dr = 2GM^2/R$ and finally
$$
E_g = {1\over 2}M\phi(R) - {GM^2\over R},~~ H = {1\over 2}M\phi(R) + 
{1\over 2}{GM^2\over R}.\eqno(4.20)
$$
So far, the only difference between our calculation and those of Eddington 
and others is the fact that we have left open the zero point of the
potential. Eddington's field is $\phi$(Eddington) =  $ -\phi- GM/R$ and
his boundary condition $\phi$(Eddington) = 0 at the surface, amounts to
$$
\phi(R) = -{GM\over R}.~~(\rm Eddington's~choice)
$$
 According to (4.16) this is the same as
$ 
\phi(0) = - (MG/\R) - cT(0).
$ 
But, as we have emphasized already, $\phi(0) = \phi[0]$ is a constant
functional,  independent of the dynamical variables. Therefore, Eddington's
choice is not only {\it ad hoc}~ but, in the context of the action principle,
wrong!  In fact, we know of no physical theory in which the manifold of
physical states is restricted to a single energy surface in phase space.
 
The total energy provided by the action principle is  given
by (IV.20) and (IV.16),
$$
H[\rho,T] =  {1\over 2}M\Big( {GM\over R} + cT(0)  +\phi[0]\Big).
$$
It depends on the value $M$ chosen for the constant of the motion,  
  the initial value $T(0)$,
and the choice of an inessential zero point for the gravitational potential.

 {\it We feel justified to conclude that the insistence on an action 
principle is much more than an aesthetic preference; it is an
essential aid to avoid fortuitous conclusions. }
  
 We have chosen to investigate the case $n = 3$, since Eddington's 
calculations are valid in that case only. As was shown, they lead to no
conclusion even in that case. The statement that the static
configurations are stable for $n < 3$ and unstable for $n\geq 3$ may be
correct, but to say that it was proved by the argument first advanced by
Perry, or by the same calculation repeated in many of our modern textbooks,
is an exaggeration. 

\bb

\ce{\bf IV.4. The polytropic gas sphere. Stability }

We use the lagrangian
$$
\L = \rho(\dot\Phi -\vec v\,^2/2 - \phi/2 + \lambda) - \R T\rho\log k +
{a\over 3}T^4,~~ k := \rho/T^n.\eqno(IV.21)
$$
Variation with respect to $T$ gives
$$
\R(\log k - n) = {4a\over 3}T^3/\rho.\eqno(IV.22)
$$
With  $n = 3$ this makes $k$ a constant, and $
\log k = 3$ when radiation is neglected. In the remainder of
this section, we set, for all values of $n$,
$$
\rho = k T^n,~~ k ~{\rm constant}.
$$
This is the usual polytropic relation used by Eddington and others, but it
is consistent with (4.21) only when $n = 3$.
 The remaining dynamical equations are
$$
-{D v\over Dt} = \phi' + cT', ~~ \dot\rho + r^{-2}( r^2\rho v)' = 0,
$$
$$
 4\pi G  \rho = r^{-2}(r^2\phi')',~~ \rho = kT^n.
$$
 \b
$\bullet $ The static solution. Eliminate $\phi$ by $\phi' = -cT'$ and change
variables, setting $r = x/\alpha$, $\alpha$ constant, Poisson's equation
becomes
$$
{4\pi G   k\over c\alpha^2} x^2T^n + (x^2T')' = 0,
$$
where the prime now stands for differentiation with repect to $x$. Set
$f(x) = T(x)/T(0)$ and $ \alpha =  \sqrt{4\pi G / cT(0) }$ so that
finallly
$$
x^2 f^n +(x^2 f')' = 0,~~ f(0) = 1,~~ f'(0) = 0.
$$
The solution decreases monotoneously to zero at x = X, this point taken
to be the surface of the star. At the outer limit  
$ 
f(x) \propto X/x-1 +o(X-x)^n. 
$ 
The integration is done easily and accurately by Mathematica, especially so
for integer values of $n$.  The radii are, 
for $ n = 2: X = 4.355, ~n = 3: X = 6.89685636197,~n = 4: X = 14.9715.$ 

$\bullet$ For the fluctuations we assume harmonic time dependence, then the
equations are
$$
-\omega^2 r^2\delta\rho =  \big(r^2\rho\,(\delta\phi' + c\delta T')\big)', ~~
\delta\rho = nkT^{n-1}\delta T,\eqno(IV.23)
$$
$$
4\pi G r^2\delta\rho = (r^2\delta\phi')'.\eqno(IV.24) 
$$
Introduce the function
$\delta
\M = r^2\delta\phi'$. Eq.s(IV.23-4) then take the form
$$
-{\omega^2\over 4\pi G} \delta \M  =   \rho\delta \M
+ r^2\rho  c\delta T'  + {\rm constant},
$$
where the constant can only be zero, and
$$
(4\pi G) r^2(nkT^{n-1}\delta T) =\delta\M',
$$ 
Elimination of $\delta T$ leads to
$$
- {\omega^2\over 4\pi G} \delta \M  =   \rho\delta \M
+ {c\over 4\pi G kn}r^2\rho \Big({\delta \M'\over x^2
T^{n-1}}\Big)'.
$$
Changing the scale as before we get
$$
-\nu^2\delta\M = f^n\delta \M + {1\over n} x^2 f^n\Big({\delta\M'\over x^2
f^{n-1}}\Big)', ~~ \nu^2 = {\omega^2\over 4\pi G k T^n(0)}.\eqno(IV.25)
$$
 
The crucial point is the choice of the correct boundary conditions, at $x =
0$ as well as the outer surface ($x = X$). At the center the solutions
take one of two forms, $1 +  C x^2 + ...$, which is unphysical,  or else
$x^3 + C x^5 + ..$. Accordingly we set
$$
\delta \M(x) = x^3 g(x),~~ g(0) = 1,~~g'(0) = 0.\eqno(IV.26) 
$$
The boundary conditions at the outer boundary
are determined by the fact that the mass is conserved, 
$$
\delta M = \delta\M(X) = 0.
$$
The equations then imply that the null point is of order $n$. With these
boundary conditions (IV.25) becomes a well defined Sturm-Liouville problem
with an essentially self adjoint, second order differential operator.

Numerical calculations with the help of Mathematica are not difficult in the
case of integer values of $n$.
  It is found that, when $n= 2$ and for $n =3$, 
$\delta\M(X)$ is positive in the whole range, for all negative values of
$\nu^2$ and for positive values below  a limit $\nu_0^2 $ that is about .06
for $n = 2$ and compatible with 0 for $n = 3$. The latter is the first,
nodeless solution of a sequence of solutions that we have not determined in
detail.  The function falls to zero at the surface,  where 
 there is an $n$th order zero. Above this lowest value of
$\nu^2$ is a discrete set of other values of
$\nu^2$ at which the boundary condition is satisfied.

At the special value $n = 3$ the `ground state', the  
lowest value of $\nu^2$, has approached very close to zero.

Polytropes with $n = 4$ are widely believed to be unstable, but 
a positive proof of this is not known.
We have searched for harmonic solutions with negative values
of $\nu^2$. The value $n = 4$ is indicated
because it is the only integer in the interesting range, and because
Mathematica is much more managable in this case.  (Accuracy is lost when
non integral powers of negative numbers appear at the end point.) 
There seems to be a discrete, decaying nodeless mode with
$\nu^2 = -.015796$, but a bifurcation at this point in
parameter space makes the conclusion uncertain. We carried the calculation
to 15 significant figures in $\nu^2$ but solutions do not
converge towards a function that vanishes at the surface. 
To overcome this difficulty we reformulated the problem in terms of the
variational calculus. The ``solution" found for $\nu^2 = -.015796$,
truncated near both ends, was used as a trial function, to show
conclusively that the spectrum of $\nu^2$ extends this far.
Among may papers on this topic we mention Cowling (1936) and Ledoux (1941).

\b\b

\ce{\bf IV.5. The case $n = 3$} 

This case is widely believed to mark the boundary between stable and
unstable polytropes. The equations are conformally invariant
and a time independent solution is  found by an infinitesimal conformal
(homology) transformation,
$$
\delta f = rf' + f.\eqno(IV.26)
$$
This does not represent an instability, but a ``flat direction", a
perturbation from which the system does not spring back, nor does it run
away. There must also be a second solution, linear in $t$, of the form 
$$
\delta f = t(rf )',~~\delta\rho = t(r\rho ' + 3\rho ).
$$
The equation of continuity  becomes
$  
  r\rho ' + 3\rho   + v\rho' + r^{-2}(r^2 v)'\rho   = 0,
$  
whence $ v = -r$. 

This linear perturbation is the first order approximation to the exact
solution found by
 Goldreich and Weber (1980), of the form
$$
f(r,t) = {1\over a(t)}\tilde f(x),~~ x = r/a(t).
$$
The continuity equation is solved by $v = \dot ax$; thus
$ \Phi = -(\dot a/a)(r^2/2) $, and
$$
\dot \phi - \vec v\,^2/2 = -a\dot a x^2/2 = cT + \phi.
$$
This leads to
$$
\tilde\phi = a(t)\phi \propto \tilde f + \kappa a^2\ddot a x^2/6,~~ \kappa
= 3k^{1/3}/c,
$$
 and Poisson's equation becomes
$$
\tilde f^3 +  {1\over x^2}(x^2\tilde f')' = {-\kappa\over x^2}a^2\ddot a
x^2/6 = - \kappa a^2\ddot a = \lambda,~~{\rm constant}.\eqno(IV.27)
$$ 
There is a first integral, 
$$
{\kappa\over 2} \dot a^2 - \lambda/a = C,~~{\rm constant}.
$$
Rescaling of $t$ and $a$ reduces this to one of three cases
$$
\dot a = \sqrt{1+1/a},~~ \dot a = \sqrt{1-1/a},~~ \dot a = 1/\sqrt a,
$$
but only the first is compatible with analyticity at $ t = 0$, thus
$$
t = \sqrt a\sqrt{1+a} - {\rm arcsinh}\sqrt a.
$$
Setting $ a = 1+b$ we find
$$
t = \sqrt{1/2}(b-b^2/2) + o(b^3)
$$
The factor $a(t)$ is  zero at a finite, negative value of $t$ and
increases monotoneously to infinity, passing through 1 at $t = 0$.
We can of course reverse the direction of flow of $t$ to get collapse in
the finite future. 

Eq.(IV.27) was solved numerically (Goldreich and Weber, 1980). The solution  
is similar to the solution of Emden's equation, just prolonged a little at
the outer end, so long as $0<\lambda< .00654376$. For larger values of
$\lambda $ the distribution does not reach zero and increases for large $r$.
For simillar studies of collapsing, isothermal spheres see Hunter (1977) and
references therein.

It is sure, therefore, that the polytrope with $n = 3$ is not stable.
Suitably erturbed, the star may expand or collapse, until the higher or lower
density causes a change in the equation of state. Among may paers on collapse
we may mention Arnett (1977), Cheng (1978), Hunter (1977) and Van Riper (1978).

\b\b
 
\no{\steptwo V. General Relativity}
 
   \ce {\bf V.1. Lorentz invariance}

The limitation to small velocities, small compared to the velocity of
light, is justified almost always, with the sole exception of the photon
gas. We shall now modify our treatment of the non relativistic gas of
massive particles to make it consistent with relativistic invariance.
 
We need a 4-dimensional velocity and an associated
velocity potential,
$$
v_\mu = \p_\mu \psi  =:\psi_\mu,~~\mu = 0,1,2,3,
$$
where $\psi$ is a scalar field. There is only one reasonable lagrangian
(Fronsdal 2007),
$$
{\cal L} = {\rho\over 2}
\big(g^{\mu\nu}\psi_{,\mu}\psi_{,\nu} - c^2) - V[\rho].\eqno(5.1)
$$
The metric is the Lorentzian $g =~ $diag$ ~(c^{-2},-1,-1,-1)$.
In the case of velocities small compared to $c$ we set
$$
\psi = c^2t +  \Phi
$$
and find to order $o(c^{-2})$  the non relativistic lagrangian (1.10).
Henceforth   $c = 1$.

We easily allow for a dynamical gravitational field by
generalizing the measure,
$$
A = \int dt d^3x \sqrt{-g}\,{\cal L}.
$$
In a weak, terrestrial gravitational field the usual approximation for the
metric is\break
$ 
g =~ {\rm diag.} ~ (1-2gz,-1,-1,-1),
$ 
which leads to (2.10).

The concept of energy (density) is all-important in thermodynamics
   and in relativistic field theories
but ill defined in General Relativity.
   However, as long as we 
limit our attention to time independent configurations, we expect to be
on relatively safe grounds when we identify the energy density with the
time-time component of the energy-momentum tensor,
$$
T_{\mu\nu} =   \rho \psi_{,\mu}\psi_{,\nu} -
g_{\mu\nu}{\cal L} .\eqno(5.2)
$$
In the non relativistic limit $T_{00}$ is our hamiltonian
   augmented with the rest mass.

The Euler-Lagrange equations include  the conservation law
$$
\p_\mu J^\mu = 0,~~ J^\mu := \sqrt{-g}g^{\mu\nu}\psi_{,\nu}.
$$
The integral $
\int\sqrt{-g}\rho d^3x$ is a constant of the motion (for appropriate
boundary conditions) and can be interpreted as mass. This is an
essential improvement over the traditional treatment. A conserved current
also permits an application to a non neutral plasma (Fronsdal 2007).
The (conserved) mass plays a central role 
in fixing the boundary conditions in the non relativistic theory; 
to retain this feature in the relativistic extension is natural. 

\b

\ce {\bf V.2. Polytropic star with radiation}

Here  we propose to try out the lagrangian (2.7) or its relativistic version for
the mixture of an ideal gas with the photon gas. In the case that the
radiation pressure is relatively unimportant there is nothing new in this,
and in the  special case that $n = 3$ the theory is identical with that of
Eddington.

In the relativistic case the action principle offers advantages even in
this particular case. Clarification of the role of mass, which is
confused or at least confusing in the traditional treatment, is an
important part of it. Another advantage is the relative ease with which one
may proceed to study mixtures.   

Variation of the action with respect to the
temperature gives the relation  (II.10) that shows a departure from the
polytropic relation $\rho = kT^n$ when $n \neq 3$. (If this last relation is
accepted, in lieu of (II.10),   
 then from this point on the equations of motion are the same
as with other methods.) The relation between Eddington's parameter
$\beta$ and $k,n$ is
$$
{1\over \beta}  =  p_{\rm tot}/ p_{\rm gas}  =   1 +
{a\over 3\R k}; 
$$
It is constant only when $n = 3$.  
In the relativistic theory, the same relations hold; Eq.(II.10) remains
valid. The equation that determines the temperature is transcendental; 
the first approximation is   $\log k/k_0 = 3$.

Applications to real stars should await the incorporation of heat flow, not
important in the case of an isolated atmosphere and of secondary
importance in the case of the earthly atmosphere, but perhaps vital for a full
understanding of the physics.

\b\b 

\no{\steptwo VI. Conclusions}

\ce{\bf VI.1. On variational principles}

Variational principles have a very high reputation in most branches of
physics; they even occupy a central position in classical thermodynamics,
see for example the authoritative treatment by Callen (1960).
An action is available for the study of laminar flows in hydrodynamics, see
e.g. Fetter and Walecka (1960), though it does not seem to have been much used.
Without the restriction to laminar flows it remains possible to
formulate an action principle (Taub 1954, Bardeen  1970, Schutz  1970), but
the proliferation of velocity potentials is confusing and
no application is known to us. Recently, variational principles have been
invoked in special situations that arise in gravitation.

In this paper we rely on an action principle formulation of the full set of
laws that govern an ideal gas, in the presence of gravity and radiation.
To keep it simple we have restricted our attention to laminar,
hydrodynamical  flows.  

It was shown that there is an action that incorporates both of Poisson's
laws as variational equations, the temperature field being treated as any
other dynamical variable. The idea of
varying the action with respect to the temperature is much in the classical
tradition. The variational equations of motion are exactly the classical
relations if radiation is neglected, or if $n = 3$. 

The first encouraging result comes with the realization that
the hamiltonian gives the correct expression for the internal
energy and the pressure, including the contributions of 
radiation, under the circumstances that are considered in classical
thermodynamics; that is, in equilibrium and in the absence of gravitation. This
is an indication that the theory is mathematically complete, requiring no
additional input from the underlying microscopic interpretation. This conclusion
is reinforced by an internal derivation of a virial theorem.

Into this framework the inclusion of a gravitational field is natural. 
Inevitably, it leads to pressure gradients and thus also temperature gradients.
If other considerations, including the heat equation, are put aside, then the
theory, as it stands, predicts the persistence of a temperature gradient in an
isolated system at equilibrium. The existence of a temperature gradient in an
isolated thermodynamical system is anathema to tradition, and further work is
required to find the way to avoid it, or to live with it. Physical
considerations indicate that the answer is to be found in the phenomenon of
convection. The theory in the present form can be applied when convection is not
important.

A secondary but satisfying result of this work has been the application of
the action principle to the study of the energy concept. Without a well
defined hamiltonian it is quite impossible to attach an operative meaning to
any expression for the value of the energy; it is always
defined up to an additive constant, independently for each solution  of the
equations of motion. With a hamiltonian at our disposal we are  in a
position to give voice to our misgivings concerning the way that ``energy"
has been invoked in some branches of physics over a period of over 100 years.
Though we conclude that past demonstrations of instabilities of polytropes
are inconclusive, we do not suggest that the results are wrong. It is of course
agreed that $n = 3$ represents an important bifurcation point. 
 
We have insisted on the role played by the mass in
fixing the boundary conditions, verified for 3 different
atmospheres. The existence of a conserved current and the associated
constant of the motion is especially important in the context
of General Relativity where the absence of this concept casts a shadow of
doubt on the choice of boundary conditions (Fronsdal 2008). Indeed it is
strange that the equation of continuity, a major pillar of nonrelativistic
hydrodynamics, has been abandoned without protest in the popular
relativistic extension. See Kippenhahn and Weigert  (1990), pages 12-13.
\b
The interaction of the ideal gas with electromagnetic fields has been
discussed in a provisional manner. The transfer of
entropy between the two gases is in accord with the usual treatment of each
system separately.
 
 \b 

\ce{\bf VI.2. Suggestions} 

(1) It is suggested that observation of the diurnal and seasonal
variations of the equation of state of the troposphere may lead to a better
understanding of the role of radiation in our atmosphere. The centrifuge
may also be a practical source of enlightenment. We understand that modern
centrifuges are capable of producing accelerations of up to 10$^6 g$.
Any positive result for the temperature gradient in an isolated gas would
certainly have important theoretical consequences.

(2)   We suggest the use of the
lagrangian (2.7), or its relativistic extension, with
$T$ treated as an independent dynamical variable and
   $n' = n$. Variation with respect to $T$
yields  the adiabatic relations between $\rho$ and
$T$, so long as the pressure of radiation is negligible, but for
higher temperatures, when radiation becomes important, the effect is to increase
the effective value of $n'$ towards the ultimate limit 3, regardless of the
adiabatic index $n$ of the gas. See in this connection the discussion by
Cox and Giuli (1968),  page 271.\break
In the case that $n = 3$ there is Eddington's treatment of the
mixture of an ideal gas with the photon gas.  But most gas spheres have a
polytropic index somewhat less than 3 and in this case 
the ratio $\beta = p_{\rm gas}/p_{\rm tot}$ may not be constant throughout
the star. The lagrangian (2.7), with $n$ identified with the 
adiabatic index of the gas, gives all the equations that are used to
describe atmospheres, so long as radiation is insignificant.
With greater radiative pressure the polytropic index of the atmosphere is
affected. It is not quite constant, but  nearly so, and it approaches
the upper limit 3 when the radiation pressure becomes dominant. Eddington's
treatment was indicated because he used Tolman's approach to relativistic
thermodynamics, where there is room for only one density and only one
pressure.   Of course, all kinds of mixtures have been studied, but the
equations that govern them do not supplement Tolman's gravitational concepts in
a satisfactory  manner, in our opinion. Be that as it may, it is patent that the
approximation  
$\beta $ = constant, in the works of Eddington and Chandrasekhar, is a device
designed  to avoid dealing with two independent gases.

\b\b 
\no{\steptwo Acknowledgements}

I thank R.J. Finkelstein, R.W. Huff and P. Ventrinelli for discussions.

\b\b 
 
\no{\steptwo References}

\no Arnett, W.D., Ap.J. Suppl., {\bf 35}, 145 (1977). Ap. J. {bf 218}, 815
(1977). 
 
\no Bardeen, J.M., A variational principle for rotating stars
in General Relativity, 

 Astrophys. J. {162}, 7 (1970).

\no Bernoulli, D., ~ Argentorat, 1738.

\no Boltzmann, L., Wissenschaftlidhe Abhandlungen, Hasenoehrl, Leipzig 1909. 

\no Callen, H.B., {\it Thermodynamics}, John Wiley N.Y. 1960. %QC311 C13t.

\no Carnot, S., quoted by Emden (1907).

\no Castor, J., {\it Radiation Hydrodynamics}, Cambridge U. press,  2004.
 
\no Chandrasekhar, S., {\it An Introduction to Stellar Structure}, U.
Chicago Press 1938.

\no Cheng, A.F., Unsteady hydrodynamics of spherical gravitational
collapse, 

Astrophys. J., {\bf 221}, 320-326 (1978).

\no Cox, J.P. and Giuli, R.T., Principles of stellar structure, Gordon and
Breach, 1968.

\no Cowling, T.G., Monthly Notices Astr. Soc. {\bf 96}, 42 (1936).

\no De Groot, S.R., {\it An Introduction to Modern Thermodynamical
Principles}, Oxford U. 

Press, 1937. % QC311 U12i.

\no Emden,  {\it Gaskugeln}, Teubner 1907.

\no Euler, H.,   {\it \"Uber die Streuung von Licht an Licht nach der
Diracschen Theorie},

 Ann.Phys.  {\bf 26} 398-?  (1936).

\no Fetter, A.L. and Walecka, J.D., {\it Theoretical Mechanics of Particles
and Continua},
 
\no Finkelstein, R.J., {\it Thermodynamics and statistical physics}, W.H.
Freeman 1969.

\no Fronsdal,  C.,  Ideal Stars and General Relativity,  

Gen.Rel.Grav. {\bf 39} 1971-2000 (2007), gr-qc/0606027.

\no Fronsdal, C., Reissner-Nordstrom and charged polytropes,

Lett.Math.Phys. {\bf 82}, 255-273 (2007).  

\no  Fronsdal, C., Stability of polytropes, Phys. Rev. D. (to appear), arXiv
0705.0774 [gr-cc].

\no Goldreich, P. and Weber, S.V., Homologously collapsing
stellar cores, 

Astrophys.J. {\bf 238} 991-997 (1980).

\no Graeff, R.W., Viewing the controversy Loschmidt-Boltzmann/Maxwell through

macroscopic measurements of the temperature gradients in vertical columns of
water, 

 preprint (2007).  

\no Holman, J.P. {\it Thermodynamics}, McGraw-Hill, N.Y. 1969.

\no Hunter, C., Collapse of unstable isothermal apheres,  Astrophys. J.,
{\bf 218} 834-845 (1977).  

\no Karplus, R. and Neuman, M.,
Non-linear Interactions between Electromagnetic Fields,

 Phys.Rev. {\bf 80} 380-385 (1950).

\no Kelvin, Thomson, W., Collected Mathematical and Physical papers, Vol. 5,
 232-235. 
\ve

\no Kelvin, Thomson, W., Collected Mathematical and Physical papers, Vol. 3,
255-260. 

Cambridge U. Press 1911.
  
\no  Kippenhahn, R. and Weigert, A, ``Stellar Structure and
Evolution", Springer-Verlag 1990.

%Page 67, ``complete, thermally adjusted equilibrium, governed by equations
%that allow 

%for a temperature gradient".

\no Lane, H.J.,   On the Theoretical Temperature of the Sun, under the
Hypothesis of a gaseous 

Mass maintaining its Volume by its internal Heat,
and depending on the laws of gases 

as known to terrestrial Experiment, Amer.J.Sci.Arts, Series 2, {\bf 4}, 57- 
(1870).

\no Ledoux, P., On the vibrational stability of gaseous stars, Astrophys.
J., {\bf 94}, 537-548 (1941).

\no Loschmidt, L., Sitzungsb. Math.-Naturw.
Klasse Kais. Akad. Wissen. {\bf 73.2} 135 (1876).  

 \no Maxwell, J.C., The London, Edinburgh and Dublin Philosophical Magazine
{\bf35} 215 (1868).

\no Mazur, P. and Mottola, E., Gravitational Vacuum Condensate Stars,

 gr-qc/0407075

\no McKenna, J. and Platzman,  P.M., Nonlinear
Interaction of Light in Vacuum,

  Phys. Rev. {\bf 129} 2354-2360 (1962).

\no Milne,E.A., Statistical Equilibrium in relation to the Photoelectric
Effect and its 

Application to the Determination of Absorption Coefficients,
Phi.Mag. {\bf 47}, 209 (1924).

\no Milne, E.A., Polytropic equilibrium 1.  the effect of configurations
 under given external 

?Handbuch Astrophys. {\bf 3} (1930). Pages 204-222.

%pressure.

\no M\"uller, I., {\it A History of Thermodynamics}, Springer, Berlin 2007. 

\no Panofsky W.K.H.  and Philips, M., {\it Classical Electricity and
Magnetism}, 

Addison-Wesley, Reading Mass.  1962.

\no Poisson, S.D., {\it Th\'eorie mathŽmatique de la chaleur},  1835.

\no Ritter, A., A series of papers in Wiedemann Annalen, now Annalen der
Physik,

 For a list see Chandrasekhar (1938). The volumes 5-20 in
Wiedemann 

Annalen appear as the volumes 241-256 in Annalen der Physik.

\no Rosseland, S., Oslo Pub., No. 1, 1931.

\no Rosseland, S. {\it Theoretical Astrophysics}, Oxford U. Press 1936.
 
\no Saha,  M.N. and Srivastava, B.N., {\it A treatise on heat}, The Indian
Press, 1935. 

\no Schutz, B.F. Jr., Perfect fluids in General Relativity: Velocity
potentials and a  variational 

principle, Phys.Rev.D {\bf 2}, 2762-2771 (1970). 

\no Schwarzschild, K., Ueber das Gleighgewicht der Sonnenatmosph\"are,

G\"ottinger Nachrichten,  41-53 (1906).

\no Stanyukovich, K.P., {\it Unsteady motion of continuous media},
Pergamon Press N. Y. 1960.
 
%\no Sommerfeld, A., {\it Thermodynamics and Statistical Mechanics}, Lectures
% Theoretical Physics,  

%  Vol. 5, Academic Press 1966, p121-134. QC311 S69tE.

\no Taub, A.H., General relativistic variational principle for
perfect fluids, 

Phys.Rev. {\bf 94}, 1468 (1954).

\no Thomson, W., Lord Kelvin, On Homer Lane's problem of a spherical
gaseous nebula, 

Nature {\bf 75} 232-235 (1907).

\no Thomson, W., Lord Kelvin, On the convective equilibrium of temperature in
the 

atmosphere, Manchester Phil.Soc. {\bf 2} , 170-176 (1862). 

\no Tolman, R.C., {\it Relativity, Thermodynamics and Cosmology},
Clarendon, Oxford 1934.

\no Tolman, R.C., The electromotive force produced in solutions by
centrifugal action, 

Phys.Chem. MIT, {\bf 59}, 121-147 (1910). 

\no Van Riper, K.A., The hydrodynamics of stellar collapse, Astrophys. J.
{\bf 221} 304-319 (1978).
  
\no Vanderslice et al, {\it Thermodynamics}, Prentice-Hall 1966.

\no Waldram, J.R., {\it The theory of electrodynamics}, Cambridge U. Press
1985. 
\ve

\end

\no Rees, M.F., Effects of very long wavelength primordial gravitational
radiation,

 Mon.Not.Astr.Soc. {\bf 154} 187-195 (1971).

\no Putterman, S. and Uhlenbeck, G.E., Thermodynamic equilibrium of Rotating
superfluids, 

Phys. Fluids, {\bf 12}, 2229-2236 (1969).

\no  Davidson, R.D., {\it Theory of Non-Neutral Plasmas}, Addison-Wesley
1990.

\no Chandrasekhar, S. and Henrich, L.S., Stellar models with isothermal
cores, 

Astrophys. J. {94}, 525-536 (1941).

\no Pipard, A.B.,  {\it Elements of Classical Thermodynamics}, Camb. U.
Press 1966.

The lagrangian (2.7) is thus   successful in accounting for the
properties of a polytropic atmosphere (not just the adiabatic atmosphere)
constituted by an ideal gas. If the gravitational force is included  it
predicts a temperature gradient that is verified experimentally.
Till now we have seen no explanation of the strange fact that
real atmospheres are polytropic. The usual interpretation is, we think, that the
temperature gradient is created by the incoming radiation. 
Emden's remark on heat flow reminds us, in the first place, of the physical
reality of heat flow. Physics demands that it be continuous. Since the gradient
of the temperature is not zero at the lower boundary heat is entering from
below and the system is not isolated.   

Let us accept that the approach to equilibrium is accompanied by heat flow from
hot to cold, at a rate that depends on the properties of the gas and on its
state,   through the heat equation. Our present lagrangian describes a
situation in which heat flow is absent from the equations of motion;  
there are no terms involving derivatives of $T$,
temporal or spacial. The absence of
$\dot T$ is natural, since we are describing stationary states.
But the heat equation   (we suppose that it is valid although
it is not incorporated into our dynamical framework) then implies that $C\vec
\btd T$ is constant. Emden suggests that this is indeed the case, and that the
heat equation is satisfied. If that is so, then we understand that the only
way that the flow enters into the dynamics of the polytropic atmosphere is
through the boundary conditions.  

{\it What we find unsatisfactory, however, is the fact that there is no clear
relationship betwee the adiabatic lagrangian and the isolated, isothermal gas.}
Ideally, a parameter should be present that would allow to reduce the intensity
of radiation to zero, ending with the isolated, isothermal atmosphere in the
limit. The parameter $k_0$ determines the temperature gradient,
but  the polytropic index is unaffected by a change of $k_0$.  

There is a concept of convection driven by the radiation,
though it is difficult to justify it in the case that
$\gamma'>\gamma$, when the atmosphere is stable to convection. But convection
must cease in the final approach of 
an isolated gas to equilibrium. The mathematical difficulty may be due, in part,
to the fact that the mass flow is not taken into account.
 
The simple answer to this is that the isolated atmosphere remains adiabatic!
That is not in contradiction with the heat equation, but it would imply a
constant heat flow that is difficult to accept. Yet, this would provide a neat 
explanation of why the polytropic index takes a value that depends strongly on
the properties of the gas but very weakly on the radiation.
 
In Section II.3 we shall try to say something about interactions with an
electric field. The problem of incorporating heat flow into the lagrangian is
left for the future.

\ce{\steptwo III. Nonstationary configurations}

\ce{\bf 3.1. }

In an isolated system (now we know that atmospheres are not usually isolated),
a temperature gradient implies change, the variables are not constant in time.
Therefore, we need to introduce the gradient of $T$ into the action.
Let us try adding a term
$$
C[\rho,T](\vec\btd T)^2/2.
$$
It looks like the contribution of kinetic energy and implies that the heat flow
possesses inertia. This will affect the equation that comes from variation of
$T$ and none of the others. It will not imply that the temperature change
according to Fourier's law, or any temporal change at all. We need to introduce
$\dot T$ into the lagrangian. 

This is difficult, for any term that is linear in $\dot T$, of the form
$f(T)\dot T$ is a time derivative and will make no contribution to the
Euler-Lagrange equations. The first order temporal; derivative $\dot T$ can only
appear in the variational equation if it appears in the lagrangian. A term like
$T\dot T$ makes no contribution to the Euler-Lagrange equations. Variation of
$T$ gives an equation that does not ressemble Fourier's heat equation. An example
of a term that gives the correct equation is
$$
S\dot T + C[\rho,T] \,\,\vec\btd S\cdot\vec\btd T .
$$
What makes this work is the fact that the additional variable $S$ does not occur
elsewhere in the lagrangian; indeed this may be the only way to get the heat
equation from the lagrangian. The only difficulty with this is that the physical
interpretation of the new field is unclear. Let us ignore that and work out the
properties of the theory obtained by adding this term to the lagrangian (2.10).

\b\b

\ce{\bf 3.3. The dielectric atmosphere}

Consider the lagrangian
$$
{\cal L} = \rho(\dot\Phi - \vec v^2/2 -gz + \lambda )
   -{\cal R}T\rho\log k +   f[T] + {c\rho\over T}\overline{F^2}/2.
 \eqno(3.1)
$$
The factor  $c\rho /T$ has the correct dependence
 on density and temperature,  to be interpreted as permittivity.
The average $\overline{F^2}$ is not determined by the average
$\overline{F_{\mu\nu}}$ of the electromagnetic field strength.  Therefore,
it is risky to identify
$c\rho /T$ with the numerical value of the permittivity of air, which is
.00054 for dry air at sea level, and we have nothing to say about the
observed electric field in the atmosphere.

The equations of motion are
$$
\lambda - gz -  \R T\big(1 + \log k)+{c\over T}\overline{F^2}/2  = 0,
\eqno(3.4)
$$
$$
\R \rho  (n-\log k)
+ {4a\over 3} T^3  - {c\rho\over T^2}\overline{F^2}/2= 0,\eqno(3.5)
$$
If we neglect the radiation term we must have
$$
\R \rho  (n-\log k)
={c\rho\over T }\overline{F^2}/2T,\eqno(3.5)
$$
An increase of 10 percent in the sea level temperature
 leads to a decrease of $\log k$ of about  .286.
The accompanying change in the density is comparatively
 less important and we find that
the required change in $\overline{F^2}$ is
$$
\delta\overline{F^2} = .286
  {(2.87\times 10^6) \times10^{-3}\times 2\times 294\over 5\times 10^{-4}}
=.965 \times 10^9.
$$
If this were due to a uniform electric field it would have a
 strength of  more than $3.1\times 10^4$ in cgs units or  3 million
$volts/cm$.\footnote*{Verify.} 

\b\b 
\ce{\bf 3.4. Heat loss and approach to equilibrium}

We believe that the loss of heat by radiation can be accounted
for by enlarging the system to include the electromagnetic field.
An approach of a closed system to equilibrium, to the extent that it
involves nothing more than a normalization of the temperature, 
may be encompassed by this, but it is clear that there are situations
where the electromagnetic field intervenes in more subtle ways, if at all.

For an example of special interest to this study consider an ideal gas in
complete isolation, in irrotational motion. Some of the energy may be
assotiated with this motion, and some with heat. Conceivably, after a
long time, the motion ceases and the kinetic energy becomes zero, while
the energy balance is assured by a rise in the temperature.

What we believe actally happens is this. In certain cases the motion does
not stop. If the vessel is a perfectly smooth torus with constant cross
section it is possible to imagine a perfectly uniform and
stationary circulating current.

\b\b

\b\b
\ce{\bf 3.5. Conclusions}

It seems that we have reached a conclusion, provisionally at least,
with respect to the influence of solar radiation on the atmosphere of the
earth. Though we are unable to describe the more or less uniform warming
of the atmosphere that certainly takess place, we are struck by the
difficulty of assigning the existence of a temperature gradient to the
incoming radiation.

This conclusion, if true, is shocking, for it contradicts a central
postulate of thermodynamics, according to which the parts any isolated
system, in eqiuibrium and in contact with each oter, must have the same
temperature. It is not thermodynamics in the narrowest sense that is being
questioned, but its application in the presence of gravity.

For the theory of atmospheres of moderate temperature this conclusion
would imply that the Lane-Ritter polytrope is an unexpectedly successful
inspiration.

Another surprising, tentative conclusion is that sources of
radiation are too weak, at least in the case of atmospheres similar to
that of the earth, to deflect the value of $k$ substantially away from
the critical value 3. It is therefore intresting to return to the
difficulty first raised in the Remark in Section 2.6. Namely, when the
action (2.10) is minimized with respect to both $\rho$ and $T$, we semm
to get too little freedom, too few equilibrium configurations.  Let us
examine this problem; thus we take the lagrangian (2.10), with $\mu$
as in (3.7), and in the notation (3.5),
$$
{\cal L}[\Phi,\rho,T] = \rho(\dot\Phi - \vec v^2/2 -gz + \lambda )
   -{\cal R}T\rho\log k   .~~ k = \rho/T^n \alpha_0.\eqno(2.10)
$$
The equations of motion are, the equation of continuity, and
$$
\dot\Phi - \vec v^2/2 -gz + \lambda = \R\rho( \log k +1),~~ \log k = n;
$$ the first from variation of $\rho$, the other from variation of $T$.
Note that this equation simply fixes what Ritter has called the
polytropic temperature.

\bb

\ve

\ce{\steptwo IV. Radiation}

The form (4.1) suggests the following action for a photon gas
($\epsilon = 1$)
$$
{\cal L}_{\rm ph} =   {\sigma\over 2}  F^2 -  W[\sigma],\eqno(4.3)
$$
where
$$
F^2 = {-1\over 2}g^{\mu\nu}g^{\alpha\beta}F_{\mu\alpha}F_{\nu\beta} =
\vec E^2 - \vec B^2.
$$
  This is not much better than a guess, but there is some justification for
it.

\b

\ce{\bf 5.2. Justification for the photon gas lagrangian }

In favor of the choice (4.3) we advance the following.

     	a) It has the same structure as (4.1).

				b) It is covariant and gauge invariant.

	   c) We have concluded  that the photon gas must be
accompanied by an electromagnetic field. This is surprising but easy to
understand. In the case of an atomic gas there is a variable that is
canonically conjugate to the density; it is the velocity or in our case
the velocity potential; it is needed to formulate dynamics. The
electromagnetic field plays the same role for the photon gas, it is a
direct analogue of the velocity of the atomic gas. In the case of
equilibrium it is constant,  at least $F^2$ is constant. The direction
is fixed by spontaneous symmetry breaking, in practice it is probably a
random variable unless the gas is polarized by an external agent. If
correct, this feature of the theory may be most significant. It predicts
a Seebeck effect (Seebeck 1828) (Saha 1935) (page 29, thermocouples), not
only in metals and in dielectrics, but in vacuum as well.

    Next, let us take $W = b\sigma^2$.   In that case there is more.

~~d) Variation with respect to the density field $\sigma$ gives
$2b \sigma = F^2 $; substitution of this into the action
converts the potential term to a non-linear term of the Born-Infeld type.
Such terms are often invoked, for example, to calculate the Casimir
effect (Roberts 1983), to explain non-linear effects observed with
lasers (McKemma and Platzman 1963),
and to promote relaxation in the photon gas. Quantum electrodynamics also
predicts the presence of such terms in the effective lagrangian, though it
predicts a correction of the type
$(F\tilde F)^2$ as well, an indication that the lagrangian (4.3) may need
refinement. According to Euler (1935) and Karplus and Neuman (1950),
the effective action is
$$
-F^2 + c_1 (F^2)^2 + c_2 (F\tilde F)^2\eqno(4.4)
$$
with $c_1 = (5/80)\alpha^2/m^4, c_2 = -(7/90)\alpha^2/m^4$, where $\alpha$
is the fine structure constant and $m$ is the mass of the electron.
More about this below.
\bb

~~e) Let us consider a situation where the magnetic field is
effectively zero. The term $\vec H^2$ in the hamiltonian density  $\vec
E^2 + \vec H^2$ is    sometimes referred to as the kinetic part of the
energy, so that there may be some justification for viewing the case that
the magnetic field vanishes as an analogue of the static case in ordinary
gas dynamics. In that case the energy density is
$$
T_t^t =  -2\sigma F^2 -p,\eqno(4.5)
$$
while the pressure, on shell, is
$$
p =-\sigma F^2 -  W.\eqno(4.6)
$$
The equation of motion makes $-\sigma F^2 = 2  W$, so that $p = W$ and
$T_t^t = 3W$,
giving the correct equation of state for a photon gas, namely
$$
T_t^t = 3p.
$$
It is granted that getting this result in the case that the magnetic
field vanishes, and only in that case, adversely affects the strength of
this argument. On the other hand, any general relation between the scalar
field $p$ and the energy density would violate Lorentz invariance.

f) Since the energy density is proportional to the fourth power of the
temperature (Stefan-Boltzmann law) we must have $\sigma \propto T^2$.
The photon density $n$ is proportional to $\sigma^{3/2}$ and thus $p
\propto n^{4/3}$. All this is satisfactory.

g) The phenomenon of sound propagation in the photon gas deserves a
section of its own.

\b

\ce{\bf  4.3. The propagation of sound}

It is believed that the photon gas is capable of transmitting sound, at a
speed of $c/ \sqrt 3$  or less. An explanation for this is the
analogy with the dynamics of sound propagation in a polytropic gas. A
static solution of Eq.s (1.1-2) is $\rho = 1,  \vec v = 0$. To
   first order in $ \rho-1$ and $\vec v$ a sound wave travelling in the
$z$ direction is described by
$$
    \dot\rho + v' = 0,~~ -\dot v  = p' = \gamma p/\rho.
$$
hence $  \ddot \rho =  \gamma (p/\rho)  \rho''$ (Saha 1935, page 95).
The square of the speed
of propagation $\gamma p/\rho = dp/d\rho$.  Here
$\rho$ is essentially the energy density and thus  by analogy,
$dp/d\rho= 1/3$ for the photon gas.

It is important to be aware of the fact that this argument is based
on thermodynamics assisted by hydrodynamics. More precisely, it is
based on an analogy with an ordinary gas for which dynamical
hydrodynamics is an established theory. In the case of photons we readily
accept that the thermodynamic aspects are understood but, as far as we
know, an independent theory of photon hydrodynamics has not been
developed. However, the result is established by a standard
thermodynamical argument (reference).

Let us consider the equation of motion associated with (4.3).
Normalization constants do not interfere, so we continue to suppress the
factor $1/16\pi$. The equations are then
$$
2b\sigma = F^2, ~~ \p^\mu  \sigma (A_{\nu,\mu} -  A_{\mu,\nu})=0.
$$
   In the gauge $A_0 = 0, ~~ $div$ \vec A = 0$ the first equation becomes
$$
2b\sigma = \sum_i \dot A_i^2 - \sum_{i,j}(\p_jA_i)(\p_jA_i).
$$
We consider the ``static" solution in which $  E_i = \dot A_i$ is
constant and of unit length,  $\sigma = 1/2b$, and
first order deviations $\d \sigma, \d A_i$ from it. In that case the first
equation gives
$$
b \d \sigma = \sum_j\dot A_j  \d \dot A_j.
$$
   The second equation reduces to $\p^\mu \sigma \p_\mu A_i = 0$ or
$$
\d \dot \sigma \dot A_i + \sigma ( {\p^2\over \p t^2}  -
\Delta)\d  A_i = 0,
$$
whence
$$
   \d  \ddot {\sigma ~}   \dot A_i + \sigma ( {\p^2\over
\p t^2}  -
\Delta)\d  \dot A_i = 0,\eqno(4.7)
$$
Projected on $\dot A_i$ (and multiplied by 2) it becomes (since $2b\sigma
= 1$ and $\sum \dot A_i\dot A_i = 1$)
$$
-2{\p^2\over \p t^2}\d  \ddot{\sigma~} = 2\sigma( {\p^2\over \p t^2}  -
\Delta)\sum_i \dot A_i\d \dot A_i=  2\sigma ({\p^2\over \p t^2}  -
\Delta )b\delta\sigma = ({\p^2\over \p t^2 }-\Delta)\d \sigma
$$
or $(3\p_t^2 - \Delta)\d \sigma = 0$. So finally the speed of the wave is
$ 1/\sqrt 3$. The direction of propagation is perpendicular to $\vec E$.

This happy result gives us some confidence in the lagrangian (3.3). With
numerical factors restored it is
$$
{\cal L}_{\rm ph} =  {\sigma\over 8\pi}  F^2 - b\sigma^2,~~ \eqno(4.8)
$$
The value of the constant $b$ unknown, so far, but since the energy
density is  $3b\sigma^2$ we know from the Stefan-Boltzmann law that
$$
3b\sigma^2 = \kappa T^4,~~ \kappa = 7.5607 10^{-15} {\rm erg/cm^3
deg K}.\eqno(4.9)
$$
\b
\ce{\bf  4.4. Refinement of the lagrangian}

We return to Euler's result (3.4) for the non linear lagrangian but
ignore the last term for the time being.

As is standard practice, the symbol
$$
F^2 =\vec  E^2 -\vec B^2
$$
stands for an energy density or for the energy in a volume of
$1cm^3$; everything expressed in cgs units. Euler's formula (1936) is
$$
{\cal L} = F^2\Big(1+ {\alpha^2\over 45\pi}({\hbar\over
m_ec})^4{F^2\over \hbar c}\Big).\eqno(4.10)
$$
All the factors are dimensionless, $\alpha$ is the fine structure
constant. Wiith $F^2$ it is
$$
{\cal L} = F^2\Big( 1 + {1\over 2.653326\times 10^6}({1\over
2.5913\times 10^{10}})^4
{F^2\over 3.1638\times 10^{-17}}\Big) = F^2\Big( 1+ {F^2\over 3.785
\times 10^{31}}\Big).
$$
The correction is of the order of unity when $F^2$ gets up to
$10^{31} erg$ or about $10^{17}kWh$, a field strength of about
$6\times {15} gauss$, as seen only in pulsars. The value of $F^2$ at
the classical edge of the electron is about $10^{14} erg$.

For this to agree with our lagrangian  we have to include the
standard Maxwell lagrangian; thus we propose to modify (2.6), setting
instead
$$
{\cal L}_{\rm ph} =  {1+\sigma\over 8\pi}  F^2 - b\sigma^2,~~ \eqno(4.11)
$$
The need to do so is clear once we identify $F$ with the
electromagnetic field strength. But we must review the evidence that
was presented in favor of (3.3), to see if it can be reconciled with
(4.11).

1. Perhaps one may be justified in speaking of 2 contributions to the
potential. The random fluctuations that are usually invoked to
explain the photon gas are normally summarized in the density
$\sigma$.
The usual treatment does not invoke a potential, and neither have we
done so up to now. But there are situations in which a global and
coherent field enters upon the scene as well, and then the ordinary
Maxwell action has to be included. It requires only a little positive
thinking to combine both in a single
electromagnetic potential. Thus one should ignore the term 1 in
$1+\sigma$ when dealing with the photon gas in the absence of a
global field.

2. If we were to replace $\sigma$ by $1+\sigma$ in the discussion of
sound propagation, in Eq.(4.7),
one would include that the speed is not $1/Ý\sqrt 3$ but that this as
the lower limit. That suggests that
this value would be found to apply in special situations only. We do
not know what the experimental situation is, except that the speed is
much less if a normal gas is present. Perhaps it will turn out that
attempts to excite a sound signal will result in light signals
instead.

3. Attention was called to the fact that the on shell value of the
lagrangian density coincides with the pressure, in hydrodynamics, and
also in Tolman's phenomenological treatment of relativistic
thermodynamics. But this leads to strange results in the context of
electromagnetism. The lagrangian,
in the case of a coherent beam of freely propagating photons,
vanishes. So, therefore, does the  of the pressure.
But the Pointing vector does not vanishes, and in the case of the
photon gas ``the pressure" is found
by placing an imaginary wall in the way of the photons and assuming
that they are reflected or absorbed by the wall. It would be more
accurate to speak of  ``the pressure on the wall"; clearly there are
two quite different concepts of pressure. It is the pressure on the
wall that is used to derive the ratio of 3 between energy density and
pressure, not "the pressure of the gas" in the sense of hydrodynamics.

In fact, a free photon by definition is not interacting with
anything. Free photons moving in a cavity, with or without an atomic
gas in it (besides the photon gas) can have no effect on the gas, or
they would not be free; in fact the quantity $F^2$ is a measure of
how much the photons are not free; that is, to what extent they
remember that they have been, or foresee that they shall be, in
interactions with the gas.

In conclusion, we  concede that  some of the arguments advanced in
favor of the original lagrangian (4.) lose some of their force as we
switch to (4. ). But the comparison with Born-Infeld theory compels
it.

\b

\ce{\bf  4.5. Reflections on the program}

This theory is intended to provide a theory of heat, through the
identification of the ``photon energy density" $3b\sigma^2$ with $aT^4$,
the Stefan-Boltzmann law. On the way to achieving that goal it is
proposed to develop a dynamical theory of the photon gas.

The dynamical variable of the photon gas certainly must include some
density, energy density or photon number  or some other density
related to one or both of these. We have seen that there are grounds for
identifying $b\sigma^2$ with the energy density in the sense
of thermodynamics. This identification may be only half right, but that is
not the problem that we wish to discuss here.. The question is whether
some additional field is needed. More precisely: is the appearance of an
electromagnetic field in this model an asset or is it an embarrasment?

The traditional treatment of ordinary gasses has served as a paradigm.
There too the density is an important dynamical variable, but the
formulation of a dynamical theory, besides the physical requirements,
are strong reasons to include the velocity field as an
additional, independent, dynamical variable. The traditional theory of
heat deals, in the first instance,  with the density of heat, a scalar
field, or equivalently with the temperature field, another scalar field.
In fact, the temperature is, sometimes, treated as a dynamical variable.
See e.g. Stayukovich 19
There is also a concept of heat flow, a vector field, but it is not an
independent dynamical variable, being related to the gradient of the
temperature. In laminar gas dynamics the velocity is the gradient of a
scalar field, but this scalar field is independent of the density field.

  From the point of view of the traditional theory of heat there is nothing
to suggest that an additional dynamical variable is needed. But in the
context in which we find ourselves the situation is quite different.
In the first place, we wish to give the theory a relativistic formulation,
and an action principle. The only relativistically invariant wave equation
for a scalar field is the Klein-Gordon equation, with some non-linear
generalizations. It is not impossible that a relativistic theory of heat
can be developed along these lines, but it would not be completely
satisfactory since it would have no apparent connection to
electromagnetism, the carrier of all heat flow.  In the second place we
need a coupling of the dynamical variables of heat to matter variables.
We know that the true, microscopic phenomenon responsible for heat
transfer is electromagnetic radiation, and we also have very definite
ideas as to the way that this radiation interacts with matter. It would seem
that this information must be taken into account, and that requires that
the electromagnetic field appear as one of the dynamical variables.
Finally, just as the velocity of flow needs to be defined to fully
describe the state of a gas, it may be thought that something besides
density may be required to describe the state of a photon gas. In
principle, a collection of photons is described by polarization and
wave number, both are lost in the amorphous concept of ``quantity of heat".

The analogy with ordinary gasses suggests an additional variable,
besides the density, and the origin of heat strongly suggests that this
variable be the electromagnetic potential.

But this brings us immediately to a difficulty, what  is
the direction of propagation and  polarization of
this radiation? Let us first suppose that the gas is polarized, all the
photons traveling in the same direction, a situation that may be
approximated in the case of radiation from a distant source.  In this case
it is natural to think that the electric field is oriented in the same
sense, and that a potential difference can be detected between two points
of the beam. An anologous effect is in fact observed in the case of heat
conduction in metals, it is the Seebeck effect (Seebeck 1821).

We come now to a characteristic feature of our model, and of similar
models, the relation $2b\sigma = -F^2$. If $F$ is the field of a set of
collimated, polarized photons, then $F^2 = 0$. Such a beam does not by
itself imply temperature. Heat would appear, not because of the presence
of free photons, but because of scattering and absorption; that is, the
fact that in reality the photons are not free. Such scattering can take
place in a normal gas and it is expected to take place in a photon gas as
well, but so far we do not have a photon gas in which to verify this.

To generate temperature there are two mechanisms: we may prepare a beam
of off shell photons, which implies the existence of a source, or we may
introduce a degree of incoherent dispersion in the polarization or the
direction of propagation. Let us consider the second possiblity first.
Dispersion will make $F^2 \neq 0$ and the relation $2b\sigma =
-F^2$ may imply a variable temperature. But this relation
depends on the degree of dispersion, not on the  amplitude of the
field, so we are not able, at this point, to see any relation between the
strength of the field and the associated temperature. On the
other hand, once $\sigma$ is not constant the photons no longer
satisfy Maxwell's equations, they are not free, so that brings
us to the first possiblity. There is boot strap operating:
If $\sigma$ is not constant then the solutions for the potential
do not satisfy $F^2 = 0$.

Finally, in order to understand the nature of the field we have to solve
the equations of motion. Only then shall we be able to decide if it is
measurable and relevant to the theory of heat.

\b

   \ce{\bf   5.6. Radiation and heat in a hot  gas}

Our aim is to calculate the heat generated in the photon gas by a hot body.
\b
1. Consider the static field generated by a charged particle at the origin
   of the coordinate system.
The potential will be assumed to have the form
$$
\vec A = 0,~ A_0 = f(r).
$$
Thus
$$
\vec E = {\rm grad}\, f = {\vec  x \over r}f' ,~~ \vec H = 0.
$$
and
$$
\sigma = -F^2 = \vec E^2 =  f'\,^2.
$$
The equation that we wish to solve is
${\rm div }(1+\sigma){\rm grad }f = 0$, or
$$
r^2(1+f'\,^2)f' = {\rm constant}.
$$
We shall make use of the function
$$
\chi: x \mapsto (1+x^2)x
$$
and the  unique, real valued,  inverse function $\chi^{-1}$. Thus
$$
f'(r) = \chi^{-1}({c\over r^2}) ,~~ c = {\rm constant}.
$$
This function falls of as $1/r^2 $ at infinity and behaves
as $1/r^{2/3}$ near the origin.

The presence of the photon gas softens the singularity.
It acts as an infrared regulator,
just like the soft photons of quantum electrodynamics. It is as if the
   charge were spread out and less concentrated at a point.

The temperature is proportional to $E_z$ and thus it falls off as $1/r^2$
   at ``large" distances. Of course, in the case of an elementary charge we
expect that this temperature is extremely low and undetectable. There is
a non-zero energy density but no flow. Since there is no magnetic field
the energy density is exactly three times the pressure. This pressure has
nothing to do with the Pointing vector that  in this case vanishes.

\b

2. Next,  let us consider a fixed dipole, consisting of a positive charge
at the origin and an opposite charge at the point $\vec d$ on the
$z$-axis.   The field equation is  a linear relation between
$(1+\sigma)F$ (not $F$) and the current, and the correct way to combine
the effect of the two sources is
$$
(1+\sigma)F =(1+\sigma_+)F_+ + (1+\sigma_-)F_-.\eqno(2.11)
$$
To first order in $d$ it is, schematically at first,
$$
(1+\sigma)F = d {\p\over \p z}(1+\sigma_+)F_+.
$$
Here $F_+$ has only one non-zero component,
   $(F_+)_{zt} = (E_+)_z =: E_{+z}$, so the total field has the same
property, hence the precise statement is that
$$
(1+\sigma)E_z = d {\p\over \p z}\big(1+{E_{+z}}^2\big)E_{+z}=
-2dc {z\over r^4}.
$$
Notice that we are not making direct use of the result obtained previously
for the fields $\vec E_\pm$ of the individual charges. It is the sources
that are being added, as expressed by (2.11). Also, since $\sigma = -F^2$,
the last equation determines both $F$ and $\sigma$. Here $\sigma = E_z^2$,
so
$$
E_z = \chi^{-1}(-2dc {z\over r^4}).
$$
Notice that angular distribution of the field of a dipole is distorted
   along with the dependence on distance. The radiation field is therefore
not exactly a pure dipole. Of course, beyond the immediate neighborhood of
the dipole the effect is extremely small.
\b

3. An atomic dipole oscillates. The emitted radiation is predicted by
   atomic physics. It consists of photons, uncorrelated bursts of
monochromatic radiation with random phases. The field emitted by a
dipole
located at the origin has an electric component as well as
a
magnetic component.

Let us suppose that the
  radiation is due to an
atomic, harmonic oscillator with fixed frequency
over a period of
several oscillation periods. Then it is the source and
not the field
that is monochromatic. The response is the field $(1+\sigma)
F$ and
it is satisfies
$$
\p^\mu(1+\sigma)F_{\mu\nu} \propto  \sin\omega
t,~~ \vec x\neq 0.
$$
Consequently,
$$
(1+\sigma)F_{\mu\nu}\propto
\sin\omega t,
$$
$$
(1+\sigma)^2\sigma^2 \propto\sin^2\omega
t,
$$
and
$$
\sigma = \chi^{-1}( c\sin\omega t),~~ c~{\rm independet~
of ~ time}.
$$
Thus a small amount of line broadening takes
place.

There is no radiating monopole, so we cannot construct the
oscillating
  dipole from a pair of oscillating charges. This makes
the calculation
more difficult and we shall be content to calculate
$\sigma $ to  the first
order of approximation. In that case we can
ignore the contribution of
$\sigma$ in the calculation of the fields,
to obtain the fields, and
finally $\sigma$, up to corrections of
order $\sigma^2$.

  There is no gauge in which $\vec A = 0$. There is
a gauge in which
  $A_0 = 0$, but the simplest construction is as
follows. Suppose $\vec A$
is paralell to the third axis, and use the
Lorentz gauge, so that (up to a
factor that represents  the dipole
moment):
$$
A_3 = {\sin\omega(t-r)\over  r}, ~~{\rm div} \vec A
= -
{\omega z\over r^2} \cos (t-r)- {z\over
r^3}\sin\omega(t-r),
$$
$$
A_0 =  {  z\over r^2}\sin\omega
(t-r)-
{z\over \omega r^3}\cos\omega(t-r)
$$
and
$$
E_1 = -
{\omega\over r}{zx\over r^2}\cos\omega(t-r)
- {3\over r^2}{zx\over
r^2}\sin\omega(t-r)
+{3\over\omega r^3}{zx\over
r^2}\cos\omega(t-r),
$$
$$
   E_3 =  {\omega\over r}\sin^2\theta
\cos\omega  (t-r)
+  {1\over   r^2}(1-3\cos^2\theta)\sin\omega(t-r)
  +{1\over \omega r^3}(3\cos^2\theta -1) \cos\omega(t-r),
   $$
   $$
  {\vec E}^2 = {\omega^2\over r^2}
\sin^2\theta\cos^2\omega(t-r)
+{2\o\over r^3}
\sin^2\theta\cos\o(t-r)
   \sin\o (t-r)
   $$
   $$
+ {1\over
r^4}(1+3\cos^2\theta) \sin^2\o (t-r)  +{2\over
r^4}
\sin^2\theta\cos^2\o (t-r).
$$
Further, the magnetic field
is
$$
H_1 = -{\omega y\over r^2}  \cos\omega(t-r)
-{y \over r^3}
\sin\omega(t-r),
$$
$$
\vec H^2 = {\omega^2\over r^2}
\sin^2\theta\cos^2\omega(t-r)
+  {2\o\over r^3}
\sin^2\theta\cos\o(t-r)
   \sin\o (t-r)  + {1\over r^4}\sin^2\theta
\sin^2\o(t-r).
$$
There is a remarkable cancellation in
$$
   \vec E^2
- \vec H^2 = {2\over r^4}\cos^2\theta
+ {2\over r^4}[\sin^2\o(t-r) -
\cos^2\o(t-r)].
$$
This implies that the field rapidly approaches a
free field as the distance from the source increases.
Averaging over
several time periods and restoring a factor $d^2$ that
represents the
dipole moment density,  we
get to leading order (Hertz 1889)
$$
E^2 +
H^2 = d^2{ \omega^2\over r^2}\sin^2\theta,~~  \vec E^2 - \vec H^2
=
d^2{2\over r^4}\cos^2\theta, ~~ \vec E\wedge\vec H =
-{\omega^2\over r^3}\vec r\sin^2\theta\cos^2\omega(t-r).\eqno(2.6)
  $$

We try to estimate both quantities in a large volume of
homogeneous gas. The energy
density gives a divergent integral, but
we cut it off at
the mean free path $r_+$ to get the energy density
(at the origin and
thus everywhere)
$$
{\cal E} = d^2\omega^2\int
drd\Omega \sin^2\theta = {8\pi\over
3}d^2\omega^2\int_0^{r_+}dr
={4\pi\over 3}d^2\omega^2r_+.
$$
We have cut off the divergent
integral at $r = r+$, where $r_+$ is the mean
free path of photons of
the given frequency in the gas.

For the integrated $\sigma$ we have
instead
$$
F^2 = 16b\sigma = 2d^2\int{drd\cos d\Omega \over
r^4}\cos^2\theta =
{8\pi\over 3}d^2\int {dr\over r^2} = {8\pi\over
3}d^2/r_-.
$$
Here we have an integral that diverges at the origin,
but we have seen
that the exact solution is less singular, so we have
cut the integral at
the effctive screening distance $r_-$.

Let us
assume, for simplicity only,  that our gas behaves like a black body,
then the
Stefan-Boltzman law tells us that
$$
{\cal E} = {4\pi\over
3}d^2\omega^2r_+ =  a T^4.\eqno(2.7)
$$
The Wien displacement
law, valid for normal temperatures, says that the
peak of the
frequency spectrum is at
$$
\omega =T/ \nu,~~\nu ~{\rm
constant}
$$
and this leads to
$$
{4\pi\over 3}{d^2\over \nu^2}r_+ =
  a T^2.
$$
We can use this result to express $\sigma$
as
$$
16b\sigma = {2\kappa \nu^2 \over r_-r_+} T^2.
$$
So
finally,
$$
\sigma = \xi T^2.\eqno(2.8)
$$

Values of the various
factor are
$$
\kappa =  7.56410^{-15} erg/cm^3\,^o\hskip-.9mm K^4
\nu = .28977/cm
$$
$$
\xi  = 1.27 \times 10^{-15}r_+r_-
\eqno(2.9)
$$
Values of the distances $r_{\pm}$ are lacking, though
it is not too hard to
give the order of magnitude of $r_+$. The
theory suggests that the product
is nearly the same for all gasses
with a high coefficient of absorption.

In the same manner we shall
calculate the Pointing vector. Averaging over time we get
$$
\vec
E\wedge \vec H = {-1\over 2} \sin^2\theta~{\omega^2\over  r^3}\vec r.
$$
This is the field at $x$ due to a dipole at the origin, aligned with
the $z$-axis. So we have a radial field
of strength ${-1\over 2} \sin^2\theta~{\omega^2\over  r^2}$.
Averaging over the orientation of the dipole
yields $(-1/3)(\omega^2/r^2)$. This is also the field at the origin
due to a dipole at $x$, hence
$$
\vec E\wedge \vec H|_0 = {1\over 3}\int{\omega^2\over r^3}\vec r d^2(x)d^3x.
$$
The main distribution comes from a regioin where we can approximate
$$
d^2(x) = d^2(0) + \vec r\cdot{\rm grad}~ d^2(0),
$$
The integral, cut off at $r = r_+$  gives
$$
\vec E\wedge \vec H|_0 = {2\pi\omega^2\over 9}r_+^2{\rm grad}d^2(0)
$$
See notes to finish.
\ve

\b

\b

\b\b
\no{\bf References}

\no Bernoulli, D., ~ Argentorat,
1738.

\no Chandrasekhar, S.

\no Euler, H.,   Über die Streuung von
Licht an Licht nach der Diracschen Theorie,

\quad Ann.Phys.  {\bf
26} 398-  (1936).

\no Fronsdal,  C.,  Ideal Stars and General
Relativity, to appear in Gravitation and

\quad General Relativity,
gr-qc/0606027.

\no  Fronsdal, C., Stability of polytropes,
arXiv
0705.0774 [gr-cc]

\no  Davidson, R.D., ``Theory of Non-Neutral
Plasmas",
Addison-Wesley
   1990.

\no Karplus, R. and Neuman, M.,
Non-linear Interactions between Electromagnetic Fields,

\quad
Phys.Rev. {\bf 80} 380-385 (1950).

\no Mazur, P. and Mottola, E.,
Gravitational Vacuum Condensate
Stars,

\no \quad
\quad~~~gr-qc/0407075

\no McKenna, J. and Platzman,  P.M., Nonlinear
Interaction of Light in Vacuum,

\quad Phys. Rev. {\bf 129}
2354-2360 (1962).

\no Milne

\no Rees, M.F., Effects of very long
wavelength primordial
gravitational radiation,

\quad
Mon.Not.astr.Soc. {\bf 154} 187-195 (1971).

\no Saha,  M.N. and
Srivastava, B.N., {\it A treatise on heat}, The Indian Press, 1935.

\no Stanyukovich, K.P., {\it Unsteady motion of continuous media},

\quad Pergamon
Press New York 1960.

\ve

\ce{\stepthree
Studies in Thermodynamics}

\no{\steptwo  I. Introduction}

1.1. Simple
hydrodynamics

1.2. Laminar flow

1.3. Variational formulation

1.4. On shell
relations and the potential

1.5. Equations of state
1.6.
The
mass

\no{\steptwo II The first law}

1.1. Thermodynamic equilibrium

1.2. The ideal gas in thermodynamics

1.3. The
ideal gas in statistical mechanics

1.4. The first law and the internal energy

1.5. The first law and the hamiltonian
\b
\no {\steptwo IV Radiation}

4.1. Lorentz invariance

Justification for the photon gas lagrangian

The
propagation of sound

Reflections on the program

Radiation and heat
in a photon gas
\b
\no {\steptwo III. A theory of temperature and
radiation}

\end{document}